\documentclass[aps,prd,twocolumn,superscriptaddress]{revtex4-1}
\usepackage{geometry}
\usepackage{graphicx}
\usepackage{epstopdf}
\usepackage{subfigure}
\usepackage{appendix}
\usepackage{amsmath}
\usepackage{indentfirst}
\usepackage{longtable}
\usepackage{bm}
\usepackage{color}
\usepackage[colorlinks,linkcolor=blue,anchorcolor=blue,citecolor=blue]{hyperref}
\usepackage{lineno}
\usepackage{float}
\begin{document}

%\preprint{}

\title{Study of the decays \boldmath$\psi(3686)\rightarrow\gamma\chi_{cJ}\rightarrow\gamma\bar{p}K^{*+}\Lambda+c.c.$ and $\psi(3686)\rightarrow\bar{p}K^{*+}\Lambda+c.c.$}


%\author{Authors list}
%\begin{small}
%\begin{center}
\author{M.~Ablikim$^{1}$, M.~N.~Achasov$^{10,d}$, P.~Adlarson$^{59}$, S. ~Ahmed$^{15}$, M.~Albrecht$^{4}$, M.~Alekseev$^{58A,58C}$, A.~Amoroso$^{58A,58C}$, F.~F.~An$^{1}$, Q.~An$^{55,43}$, Y.~Bai$^{42}$
% ...
\\
\vspace{0.2cm}
(BESIII Collaboration)\\
\vspace{0.2cm} {\it
$^{1}$ Institute of High Energy Physics, Beijing 100049, People's Republic of China\\
% ...
$^{4}$ Bochum Ruhr-University, D-44780 Bochum, Germany\\
% ...
$^{10}$ G.I. Budker Institute of Nuclear Physics SB RAS (BINP), Novosibirsk 630090, Russia\\
% ...
$^{15}$ Helmholtz Institute Mainz, Johann-Joachim-Becher-Weg 45, D-55099 Mainz, Germany\\
% ...
$^{42}$ Southeast University, Nanjing 211100, People's Republic of China\\
$^{43}$ State Key Laboratory of Particle Detection and Electronics, Beijing 100049, Hefei 230026, People's Republic of China\\
% ...
$^{55}$ University of Science and Technology of China, Hefei 230026, People's Republic of China\\
% ...
$^{58}$ (A)University of Turin, I-10125, Turin, Italy; (B)University of Eastern Piedmont, I-15121, Alessandria, Italy; (C)INFN, I-10125, Turin, Italy\\
$^{59}$ Uppsala University, Box 516, SE-75120 Uppsala, Sweden\\
% ...
$^{d}$ Also at the Novosibirsk State University, Novosibirsk, 630090, Russia\\
% ...
}}
%\end{center}

\vspace{0.4cm}
%\end{small}

%\thanks{}
%\affiliation{}


\date{\today}

\begin{abstract}
  Based on the data sample of $448.1\times10^{6}$ $\psi(3686)$ events collected with the BESIII detector at BEPCII, we present a study of the decays $\psi(3686)\rightarrow\gamma\chi_{cJ}\rightarrow\gamma\bar{p}K^{*+}\Lambda+c.c.$ and $\psi(3686)\rightarrow\bar{p}K^{*+}\Lambda+c.c.$. The branching fractions of $\chi_{cJ}\rightarrow\bar{p}K^{*+}\Lambda+c.c.$ ($J$=0, 1, 2) are measured to be $(4.8\pm0.7\pm0.5)\times10^{-4}$, $(5.0\pm0.5\pm0.4)\times10^{-4}$, and $(8.2\pm0.9\pm0.7)\times10^{-4}$, respectively, where the first uncertainties are statistical and the second systematic. The branching fraction of $\psi(3686)\rightarrow\bar{p}K^{*+}\Lambda+c.c.$ is measured to be $(6.3\pm0.5\pm0.5)\times10^{-5}$.  All these decay modes are observed for the first time.
\end{abstract}

%\keywords{}

\maketitle

% References should be done using the  \cite, \ref, and \label commands
\section{\label{s1}Introduction}
 The quark model provides a good description of both the ground states and some excited states of baryons. However, several resonances that are predicted by this model have not yet been observed, and hence there is an intense experimental effort underway to find these missing states~\cite{P1}. The baryon coupling in conventional production channels ({\it e.g.} $\gamma$-nucleon) can be quite small, but the coupling between baryons and $\chi_{cJ}$ decays via $gg$ gluons could be larger ({\it e.g.} $\psi$ or $\chi_{cJ}$ decays). For this reason, charmonium decay is a promising process to study excited nucleons and hyperons~\cite{P2}.\par

The BES Collaboration has reported a study of $J/\psi\rightarrow\bar{p}K^{+}\Lambda+c.c.$ and $\psi(3686)\rightarrow\bar{p}K^{+}\Lambda+c.c.$ decays~\cite{P3}, in which a threshold enhancement in the $\bar{p}\Lambda$ mass spectrum was observed. Throughout this paper, the inclusion of charge conjugate channels is implied. The BESIII Collaboration also reported a study of $\psi(3686)\rightarrow\gamma\bar{p}K^{+}\Lambda$ \cite{P4}, where a near threshold enhancement in the mass spectrum of $\bar{p}\Lambda$ was observed in $\chi_{c0}$ decay. This enhancement may be interpreted as a quasibound dibaryon state, or as an enhancement due to final-state interaction, or simply as an interference effect of  high-mass $N^{*}$ and $\Lambda^{*}$ states~\cite{P4}. The study of the resonant structures in the similar decay modes $\psi(3686)\rightarrow\gamma\chi_{cJ}\rightarrow\gamma\bar{p}K^{*+}\Lambda$ and $\psi(3686)\rightarrow\bar{p}K^{*+}\Lambda$ may help in the understanding of the $\bar{p}\Lambda$ threshold structure. \par

 Until now, no experimental results exist concerning the decays $\psi(3686)\rightarrow\gamma\chi_{cJ}\rightarrow\gamma\bar{p}K^{*+}\Lambda$ and $\psi(3686)\rightarrow\bar{p}K^{*+}\Lambda$. In this analysis, the branching fractions (BFs) of $\chi_{cJ}\rightarrow\bar{p}K^{*+}\Lambda$ ($J$ = 0, 1, 2) and $\psi(3686)\rightarrow\bar{p}K^{*+}\Lambda$ are measured for the first time with a data sample of $448.1\times10^{6}$ $\psi(3686)$ events \cite{P5}. Moreover, possible substructures in invariant mass spectra of $\bar{p}K^{*+}$, $K^{*+}\Lambda$, and $\bar{p}\Lambda$ are investigated.

\section{\label{s2}BESIII DETECTOR AND MONTE CARLO SIMULATION}
   The Beijing Electron Positron Collider II (BEPCII) is  a double-ring $e^{+}e^{-}$ collider running at center-of-mass energy ranging from 2.0 to $4.6~\rm{GeV}$. The BESIII detector \cite{P6} at BEPCII, with a geometrical acceptance of $93\%$ of the $4\pi$ solid angle, operates in a magnetic filed of 1.0 T provided by a superconducting solenoid magnet. The detector is composed of a helium-based main drift chamber (MDC), a plastic-scintillator time-of-flight (TOF) system, a CsI(Tl) electromagnetic calorimeter (EMC) and a resistive plate chambers (RPC)-based muon chamber (MUC). The spatial resolution of the MDC is better than 130 $\mu$m, the charged track momentum resolution is $0.5\%$ at 1 GeV/$c$, and the energy-loss ($dE/dx$) resolution is better than $6\%$ for electrons from Bhabha events. The time resolution of the TOF is 80 ps (110 ps) in the barrel (endcaps. The energy resolution of the EMC at 1.0 GeV is $2.5\%$ ($5\%$) in the barrel (endcaps). The position resolution in the MUC is better than 2 cm.\par

Simulated Monte Carlo (MC) events are used to determine the detection efficiency, optimize selection criteria and estimate the level of contamination from background processes. The {\sc geant}{\footnotesize 4}-based~\cite{P7} simulation package {\sc boost} includes a geometric and material description of the BESIII detector, detector response, and digitization models, and also tracks the running conditions and performance of the detector. The production of $\psi(3686)$ events is simulated with {\sc kkmc} \cite{P8}, where the known decay modes are generated by {\sc evtgen}~\cite{P9,P13} with their BFs taken from the Particle Date Group (PDG) \cite{P10}, and the remaining unknown decays are generated by {\sc lundcharm} \cite{P11}. Exclusive MC samples of $\psi(3686)\rightarrow\gamma\chi_{cJ}\rightarrow\gamma\bar{p}K^{*+}\Lambda$ and $\psi(3686)\rightarrow\bar{p}K^{*+}\Lambda$ are generated to determine detection efficiencies. In the signal MC simulation, the angular distribution of the decay $\psi(3686)\rightarrow\gamma\chi_{cJ}$ has the form $1+\alpha \cos^{2}\theta$ with $\alpha$=1, $-1/3$, 1/13 for $J=$0, 1, 2, respectively, where $\theta$ is the photon polar angle \cite{P12}. The weak decay of $\Lambda$ is generated with a model that includes parity violation. Other relevant decays  are generated with {\sc besevtgen} \cite{P13} with a uniform distribution in phase space.\par

\section{\label{s3}Analysis of $\bm{\psi(3686)\rightarrow\gamma\chi_{cJ}\rightarrow\gamma\bar{p}K^{*+}\Lambda}$}
\subsection{\label{s3.1}Event selection}
    The process $\psi(3686)\rightarrow\gamma\chi_{cJ}\rightarrow\gamma\bar{p}K^{*+}\Lambda$ is reconstructed with $\Lambda\rightarrow p\pi^{-}$, $K^{*+}\rightarrow K^{+}\pi^{0}$, and $\pi^{0}\rightarrow\gamma\gamma$. Events are required to have at least two positive and two negative charged tracks. For each charged track, the polar angle in the MDC must satisfy $|\cos\theta|< 0.93$. The combined TOF and $dE/dx$ information is used to form particle identification (PID) confidence levels for pion, kaon and proton hypotheses. Each track is assigned to the particle hypothesis with the highest confidence level. The identified $\bar{p}$ and $K^{+}$ candidates are further required to have their point of closest approach to the interaction point (IP) within $\pm$1 cm in the plane perpendicular to beam direction and within $\pm$10 cm in the plane of the beam direction. A common vertex constraint is applied to all $p\pi^{-}$ pairs assumed to arise from  a $\Lambda$ decay, and the production of the $\Lambda$ candidates is constrained to be at the interaction point. Only $dE/dx$ information is used for the PID of $p$ and $\pi^{-}$ candidates in $\Lambda$ decays, because many of these particles do not reach the TOF on account of their low momentum.
%The identified $p$ and $\pi^{-}$ are not required to originate from the IP.

    Photon candidates are required to have energy deposition greater than 25 MeV in the barrel EMC ($|\cos\theta|<0.8$) and 50 MeV in the end cap EMC ($0.86<|\cos\theta|<0.92$). To exclude showers from charged tracks, the angle between the direction of the photon and the nearest charged track is required to be greater than $5^{\circ}$. In addition, the angle between the direction of the photon and anti-proton is required to be greater than $10^{\circ}$ to suppress background from anti-proton annihilation in the detector.
 The measured EMC time is required to be within 0 and 700 ns of start time of the event to suppress electronic noise and any energy deposition unrelated to the event.\par

    To improve the mass resolution, the selected photons, anti-proton, kaon, and $\Lambda$ candidate are subjected to a five-constraint (5C) kinematic fit under the hypothesis of $\psi(3686)\rightarrow\gamma\bar{p}K^{+}\pi^{0}\Lambda$ with the invariant mass of the two photons being constrained to the $\pi^{0}$ mass. The $\chi^{2}$ of the 5C fit is required to be less than 70. For events with more than one combination satisfying this requirement, only the combination with the smallest $\chi^{2}$ is accepted. To veto background events from $\psi(3686)\rightarrow\bar{p}K^{+}\pi^{0}\Lambda$ and $\psi(3686)\rightarrow\gamma\bar{p}K^{+}\Lambda$, an alternative 5C (4C) kinematic fit is performed under the hypotheses of $\psi(3686)\rightarrow\bar{p}K^{+}\pi^{0}\Lambda$ ($\gamma\bar{p}K^{+}\Lambda$). We further require the confidence level of the kinematic fit for the $\psi(3686)\rightarrow\bar{p}K^{+}\pi^{0}\Lambda$ assignment to be  larger than those for the $\psi(3686)\rightarrow\gamma\bar{p}K^{+}\pi^{0}\Lambda$ and $\psi(3686)\rightarrow\gamma\bar{p}K^{+}\Lambda$ hypotheses.\par

\begin{figure}{}
  \centering
  \subfigure{
  %\begin{minipage}{0.22\textwidth}
   \centering
    \includegraphics[width=1.5in]{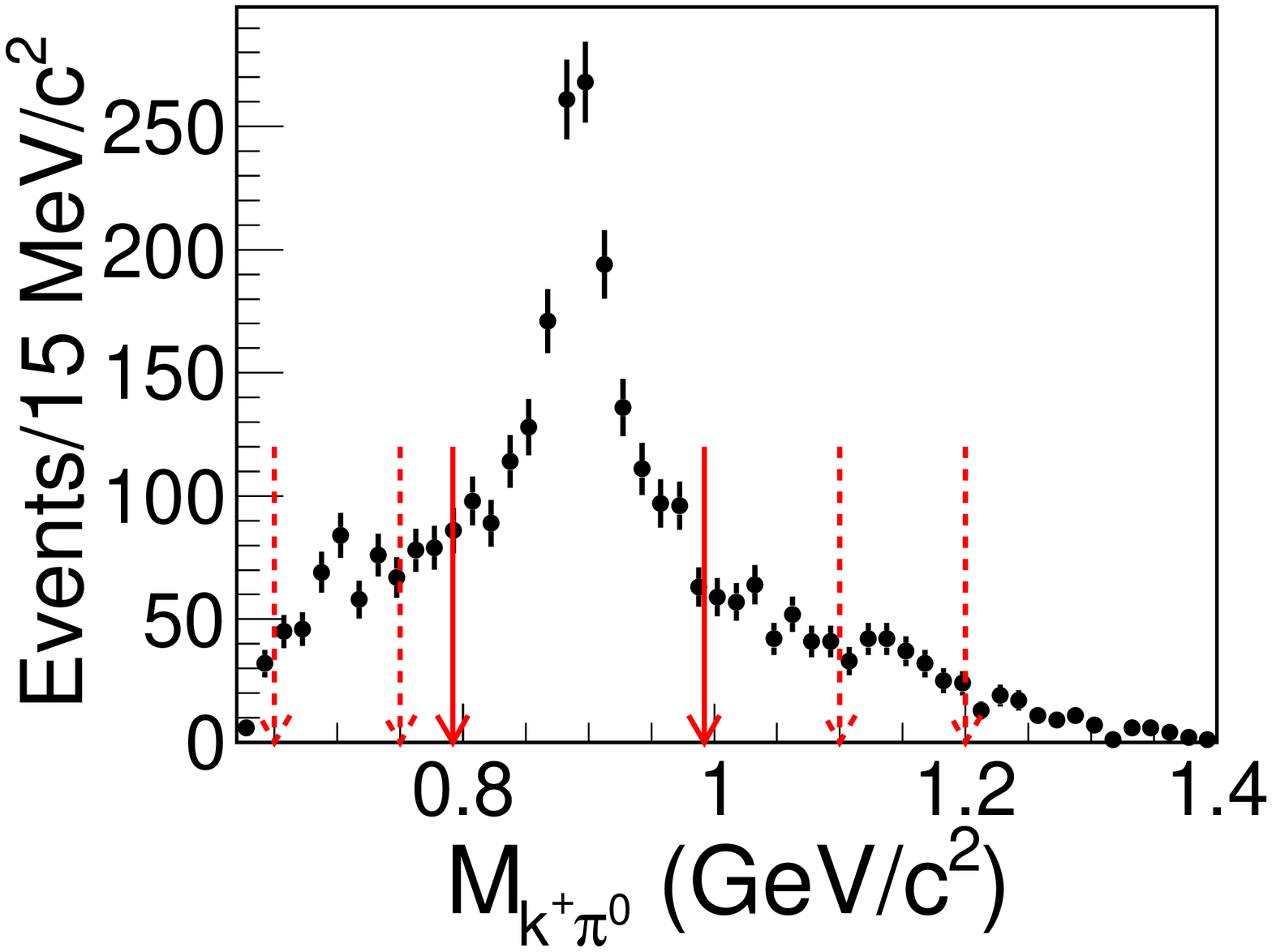}
    \put(-30,60){(a)}
  %\end{minipage}
  }
  \subfigure{
  %\begin{minipage}{0.22\textwidth}
    \centering
    \includegraphics[width=1.5in]{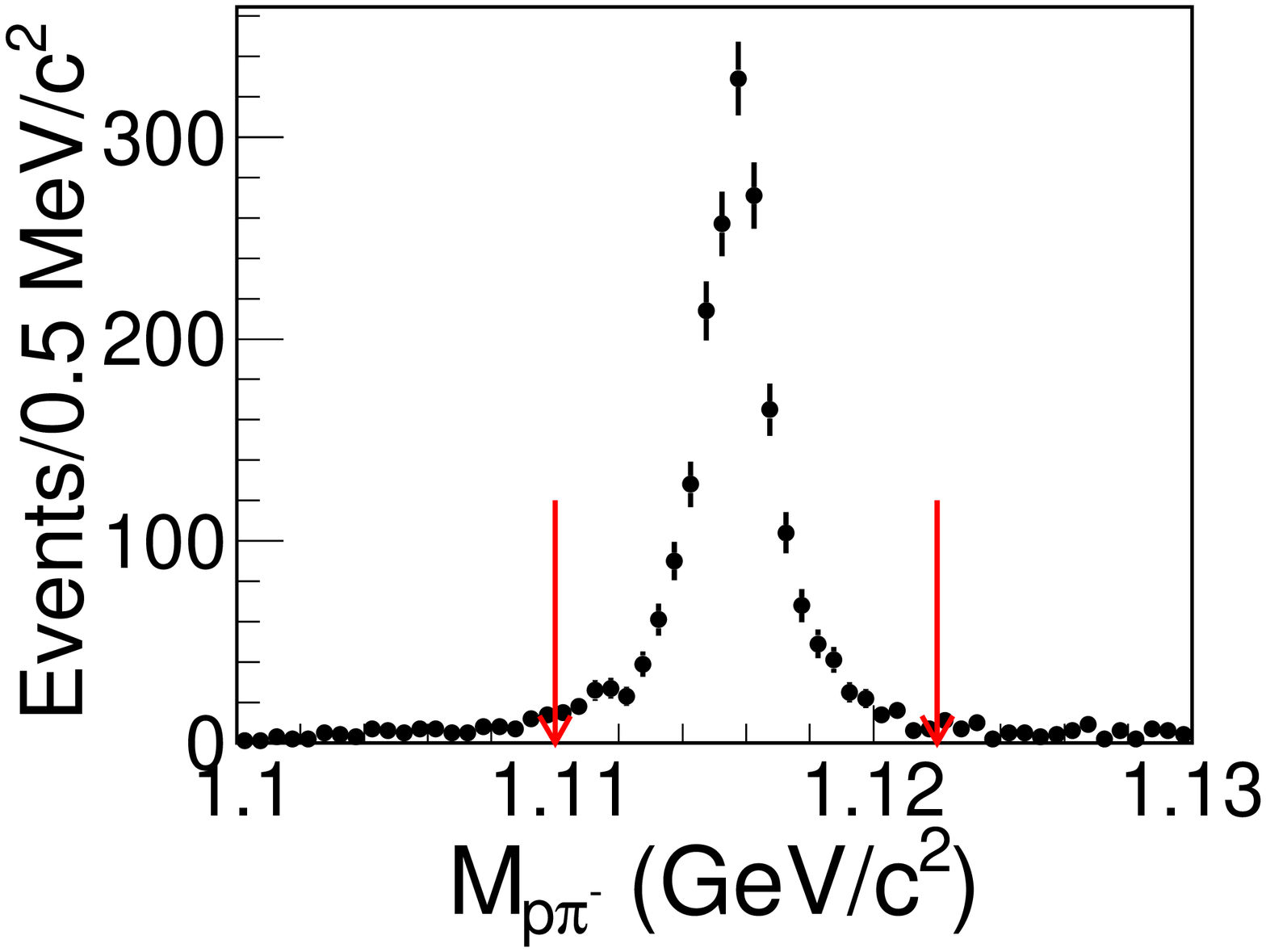}
    \put(-30,60){(b)}
  %\end{minipage}
  }
  \caption{Invariant mass distribution of (a) $K^{+}\pi^{0}$ and (b) $p\pi^{-}$. The solid arrows indicate the mass windows used as the selection criteria in the analysis. The dashed arrows indicate the sidebands region.}\label{fig2}
\end{figure}

    The  $K^{+}\pi^{0}$ invariant mass distribution is shown in Fig.\ \ref{fig2}(a), where an obvious $K^{*+}$ structure can be seen. The $K^{*+}$ candidates are selected by requiring $|M_{K^{+}\pi^{0}}-M_{K^{*+}}|<0.1~{\rm GeV}/c^{2}$, where $M_{K^{*+}}$ is the nominal mass of the $K^{*+}$ meson \cite{P10}. The $K^{*+}$ sidebands, also indicated in Fig.\ \ref{fig2}(a), are chosen to be $1.1 <M_{K^{+}\pi^{0}}<1.2~{\rm GeV}/c^{2}$ and $0.65 <M_{K^{+}\pi^{0}}<0.75~{\rm GeV}/c^{2}$. Figure\ \ref{fig2}(b) shows  the $M_{p\pi^{-}}$ distribution, from which $\Lambda$ candidates are selected by requiring $|M_{p\pi^{-}}-M_{\Lambda}|<6~{\rm MeV}/c^{2}$, where $M_{\Lambda}$ is the nominal $\Lambda$ mass \cite{P10}. Background events from $\psi(3686)\rightarrow J/\psi\pi^{0}\pi^{0}$, $J/\psi\rightarrow\bar{p}K^{+}\Lambda$ are rejected by requiring $|M_{\bar{p}K^{+}\Lambda}-M_{J/\psi}|>0.05~{\rm GeV}/c^{2}$, where $M_{J/\psi}$ is the nominal $J/\psi$ mass \cite{P10}. To remove the background from the cascade decay $\psi(3686)\rightarrow \bar{p}K^{+}\Sigma^{0}$, $\Sigma^{0}\rightarrow\gamma\Lambda$, the additional selection requirement $M_{\gamma\Lambda}>1.21$~GeV/$c^{2}$ is applied.\par

\begin{figure}
  \centering
  \includegraphics[width=0.3\textwidth]{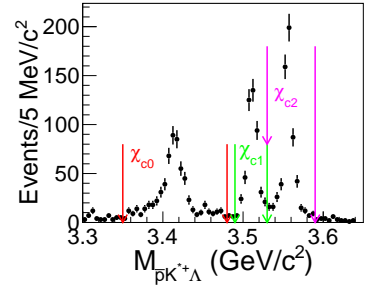}\\
  \caption{Invariant mass spectrum of $\bar{p}K^{*+}\Lambda$. The three arrow-pairs indicate, from left to right, the mass windows for $\chi_{c0}$, $\chi_{c1}$, and $\chi_{c2}$, respectively.}\label{fig3}
\end{figure}

\begin{figure*}{}
  \centering
  \subfigure{
  %\begin{minipage}{0.3\textwidth}
    \centering
    \includegraphics[width=2.0in]{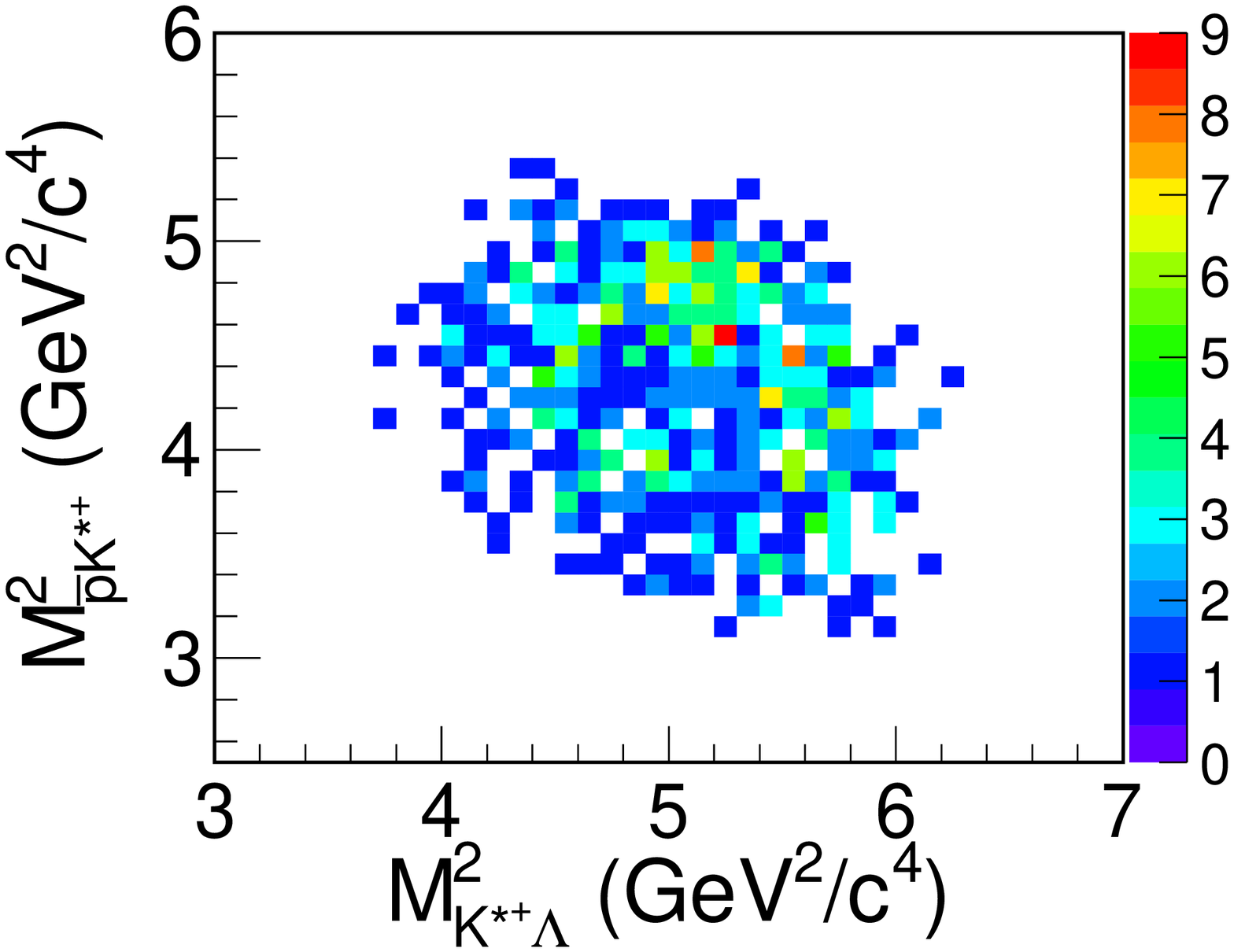}
    \put(-35,100){(a)}
  %\end{minipage}
  }
  \subfigure{
  %\begin{minipage}{0.3\textwidth}
    \centering
    \includegraphics[width=2.0in]{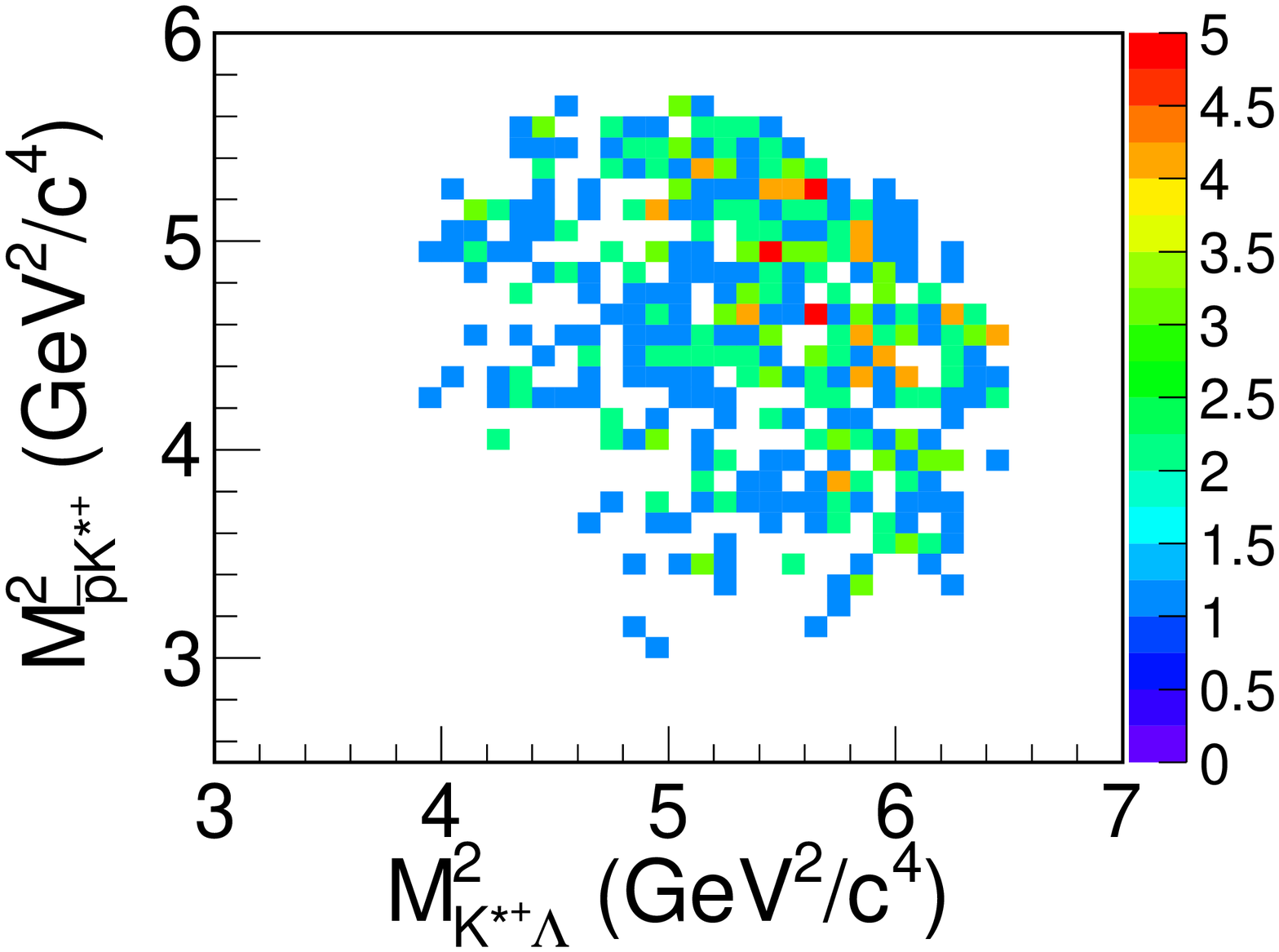}
    \put(-35,100){(b)}
  %\end{minipage}
  }
  \subfigure{
  %\begin{minipage}{0.3\textwidth}
    \centering
    \includegraphics[width=2.0in]{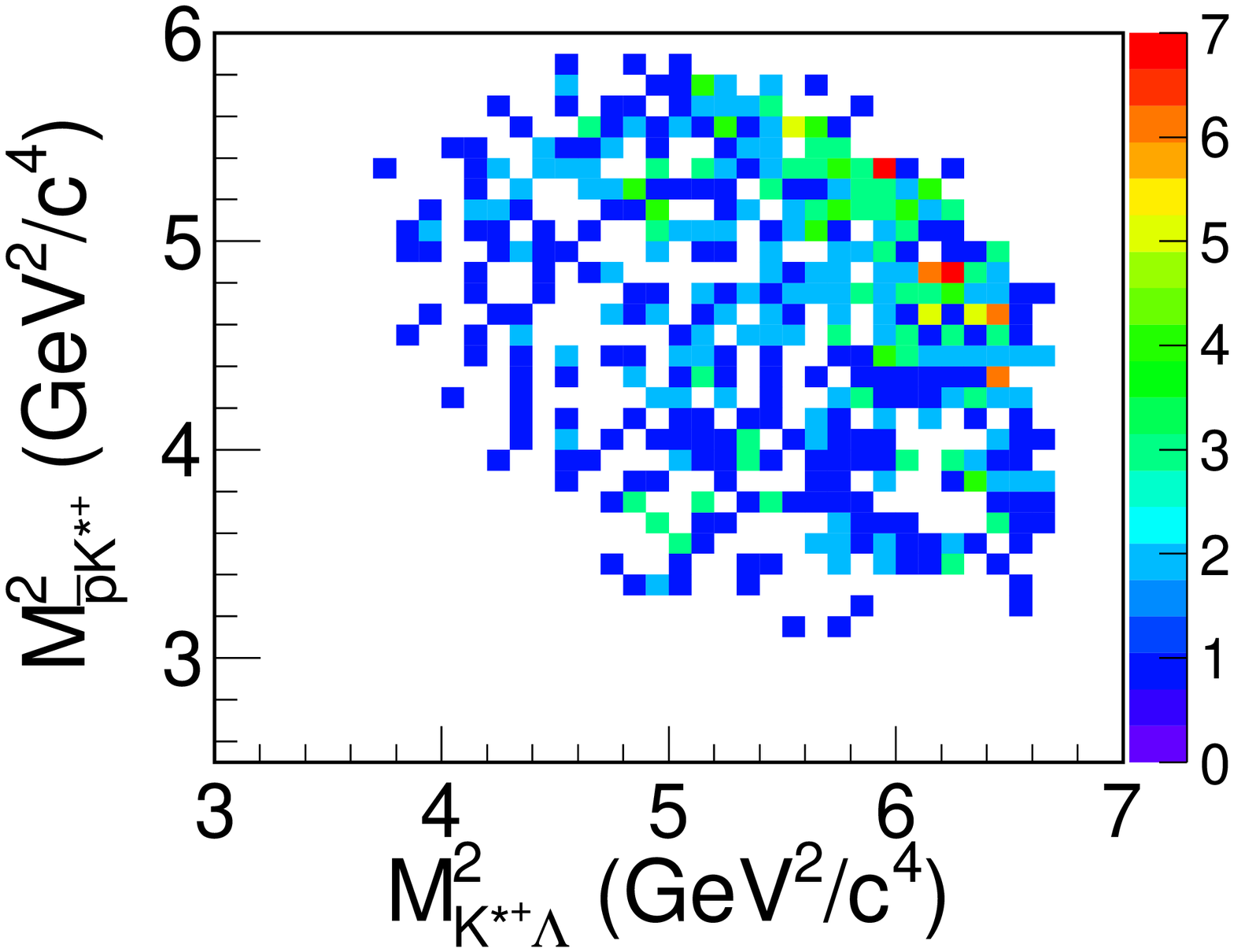}
    \put(-35,100){(c)}
  %\end{minipage}
  }
  \caption{The Dalitz plots of $\bar{p}K^{*+}\Lambda$ for $\chi_{c0}$ (a), $\chi_{c1}$ (b), and $\chi_{c2}$ (c).}\label{fig4}
\end{figure*}

    After applying these requirements, $\chi_{cJ}$ signals are clearly seen in the invariant mass spectrum of $\bar{p}K^{*+}\Lambda$, as shown in Fig.~\ref{fig3}. The mass windows used to select the $\chi_{c0}$, $\chi_{c1}$, $\chi_{c2}$ candidates correspond to about three times the $\chi_{cJ}$ width convolved with the mass resolution, which are 3.35-3.48, 3.49-3.53, and 3.53-3.59~GeV/$c^{2}$, respectively. The invariant mass spectra of the $\bar{p}K^{*+}$, $\bar{p}\Lambda$, and $K^{*+}\Lambda$ combinations and the corresponding Dalitz plots are shown in Fig.~\ref{fig4} for each $\chi_{cJ}$ state.
%No obvious substructure is evident.
No significant substructure is seen in the Dalitz plots of $\bar{p}K^{*+}\Lambda$ distributions. In order to search for the near-threshold structure of $M_{\bar{p}\Lambda}$ observed in Ref.\ \cite{P4} in the decay $\chi_{c0}\rightarrow\bar{p}K^{+}\Lambda$,
fits are performed on $M_{\bar{p}\Lambda}$ where the structure is described by a weighted Breit-Wigner resonance with parameters fixed to those reported in Ref.\ \cite{P4}. These fits return a statistical
significance for the structure of  2.1$\sigma$, 2.5$\sigma$, and 1.9$\sigma$ for the $\chi_{c0}$,  $\chi_{c1}$, and $\chi_{c2}$ states, respectively. \par


\subsection{\label{s3.2}Background study}
Using an inclusive MC sample of $506\times10^{6}$ $\psi(3686)$ events, the background from fake $\Lambda$ is found together with fake $K^{*+}$. So, the background can be categorized into the following four types: (1) events with  a genuine $K^{*+}$ and a fake $\chi_{cJ}$ ($K^{*}$, non-$\chi_{cJ}$); (2) events with a genuine $\chi_{cJ}$ and a fake $K^{*}$ ($\chi_{cJ}$, non-$K^{*}$); (3) events with fake $K^{*}$ and $\chi_{cJ}$  candidates (non-$K^{*}$, non-$\chi_{cJ}$); (4) events containing a genuine $K^{*+}$ and  a genuine $\chi_{cJ}$  ($K^{*}$, $\chi_{cJ}$). The contributions from the first three categories can be estimated by performing a two-dimensional (2-D) fit to the distribution of $M_{K^{+}\pi^{0}}$ versus $M_{\bar{p}K^{*+}\Lambda}$. The fourth type of background events come mainly from the processes $\psi(3686)\rightarrow\gamma\chi_{cJ}\rightarrow\gamma\bar{p}K^{*+}\Lambda\rightarrow\gamma\gamma\bar{p}K^{+}\Lambda$, $\psi(3686)\rightarrow\gamma\chi_{cJ}\rightarrow\gamma\bar{p}K^{*+}\Lambda\rightarrow\gamma\bar{p}K^{*+}\gamma p\pi^{-}$, $\psi(3686)\rightarrow\gamma\chi_{cJ}\rightarrow\gamma\gamma J/\psi\rightarrow\gamma\gamma\bar{p}K^{*+}\Lambda$ and $\psi(3686) \rightarrow\gamma\chi_{cJ}\rightarrow\gamma\bar{p}K^{*+}\Sigma^{0}$.
The first two of these contributions are negligible, on account of the low BF of radiative $K^{*+}$ and $\Lambda$ decays. The level of contamination coming from the other two modes is assessed by applying the selection to samples of exclusive MC events.
%Since the relative rate of radidative decays $\frac{\mathcal{B}(K^{*+}\rightarrow K^{+}\gamma)}{\mathcal{B}(K^{*+}\rightarrow K^{+}\pi^{0})}\approx10^{-3}$ and $\frac{\mathcal{B}(\Lambda \rightarrow\gamma p\pi^{-})}{\mathcal{B}(\Lambda\rightarrow p\pi^{-})}\approx10^{-3}$ \cite{P10}, the number of background events from $\psi(3686)\rightarrow\gamma\chi_{cJ}\rightarrow\gamma\bar{p}K^{*+}\Lambda,K^{*+}\rightarrow K^{+}\gamma$ and $\psi(3686)\rightarrow\gamma\chi_{cJ}\rightarrow\gamma\bar{p}K^{*+}\Lambda,\Lambda \rightarrow\gamma p\pi^{-}$ is negligible.
 For the normalization procedure, the BF of $\psi(3686)\rightarrow\gamma\chi_{cJ},\chi_{cJ}\rightarrow\gamma J/\psi,J/\psi\rightarrow\bar{p}K^{*+}\Lambda$ is estimated to be less than $10^{-5}$, which implies negligible background of less than one event from this source.
%
% GW: dont need the below information
%according to the $\mathcal{B}(\psi(3686)\rightarrow\gamma\chi_{cJ})$ \cite{P10}, $\mathcal{B}(\chi_{cJ}\rightarrow\gamma J/\psi)$ \cite{P10} and $\mathcal{B}(\psi(3686)\rightarrow\bar{p}K^{*+}\Lambda)$ measured in this analysis.
%
% GW: suppress the below
%{\color{red} The $\mathcal{B}(\psi(3686)\rightarrow\gamma\chi_{cJ},\chi_{cJ}\rightarrow\bar{p}K^{*+}\Sigma^{0})$ are approximately measured to be $(1.2\pm0.4)\times10^{-5}$, $(0.5\pm0.2)\times10^{-5}$ and $(0.6\pm0.3)\times10^{-6}$ for $\chi_{cJ}$ ($J$=0, 1, 2) {\it what does 'approximately measured' mean?  do we really want to give these numbers?}}.
%
% GW: below information moved a few sentences earlier
%The normalized number of $\psi(3686)\rightarrow\gamma\chi_{cJ},\chi_{cJ}\rightarrow\gamma %J/\psi,J/\psi\rightarrow\bar{p}K^{*+}\Lambda$ background event is less than one and negligible.
%
The normalized number of $\psi(3686)\rightarrow\gamma\chi_{cJ},\chi_{cJ}\rightarrow\bar{p}K^{*+}\Sigma^{0}$ background events is estimated to be 11.7$\pm$3.5, 5.1$\pm$2.3, 4.8$\pm$2.6 for $\chi_{cJ}$ ($J$=0, 1, 2),
where the relative BFs used to calculate these yields are estimated from dedicated studies with the same data sample.   \par

      To investigate possible background from continuum processes, the same selection criteria are applied to a data sample of 2.93 fb$^{-1}$ \cite{P15} collected at $\sqrt{s}=3.773$ GeV. After normalizing to the integrated luminosity of the $\psi(3686)$ data sample, 20.1$\pm$4.1 events survive and no peak is found in the mass spectrum of $M_{\bar{p}K^{*+}\Lambda}$. As a cross check the selection is also performed on a data sample of 44.5 pb$^{-1}$ collected at $\sqrt s=3.65$ GeV.  Only one event survives, which corresponds to 14 events when normalized to the integrated luminosity of the $\psi(3686)$ data sample, and is consistent with the result of the first study.   In the BF measurement any continuum contribution is included in the other sources of non-peaking background and the total is estimated by the 2-D fit described below.

\subsection{\label{s3.3}Branching fraction measurement of $\bm{\chi_{cJ}\rightarrow\bar{p}K^{*+}\Lambda}$}
\begin{figure*}
  \centering
  \subfigure{
  %\begin{minipage}{0.22\textwidth}
    \centering
    \includegraphics[width=2.2in]{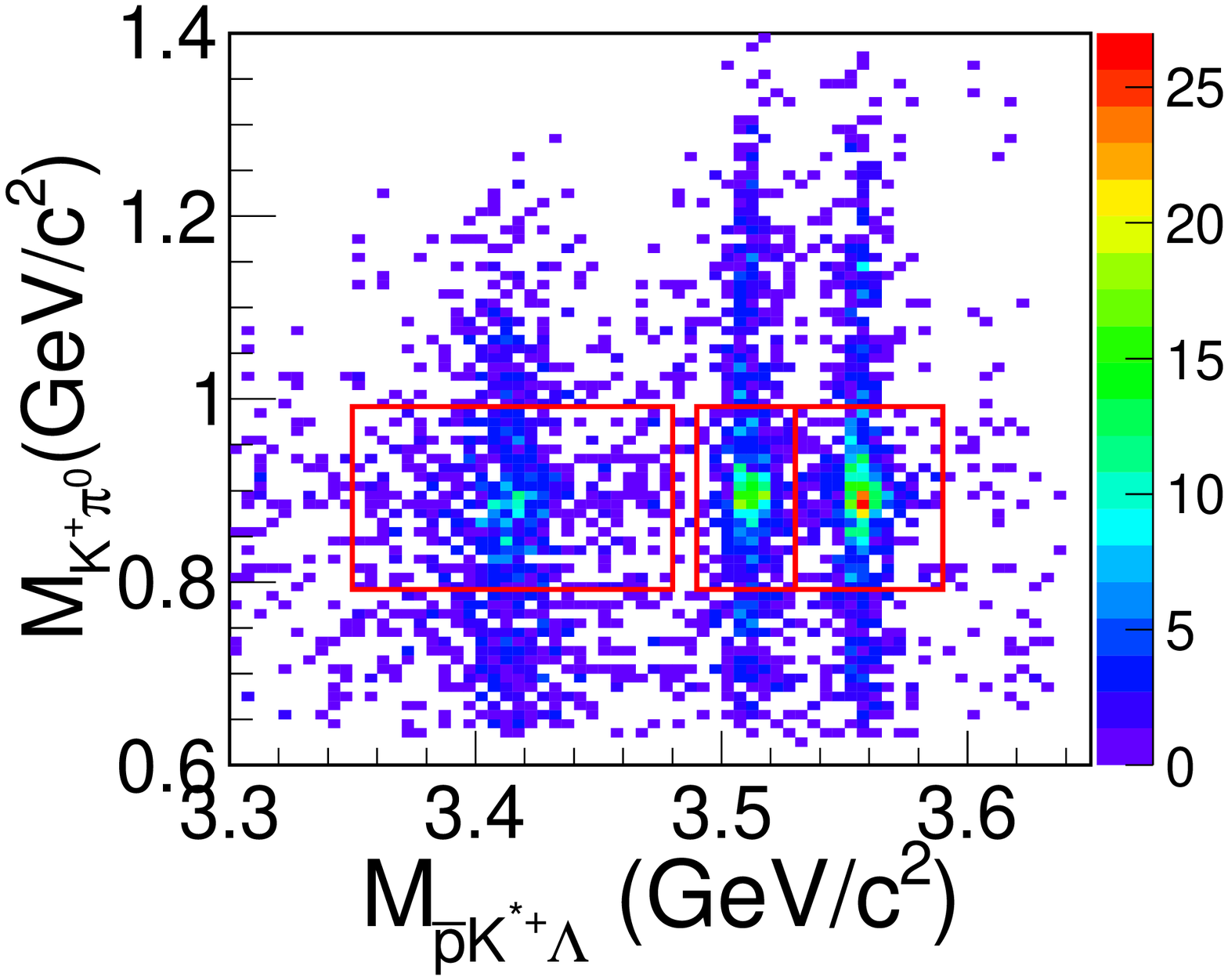}
    \put(-80,70){(a)}
  %\end{minipage}
  }\hspace{-0.08in}
  \subfigure{
  %\begin{minipage}{0.22\textwidth}
    \centering
    \includegraphics[width=2.2in]{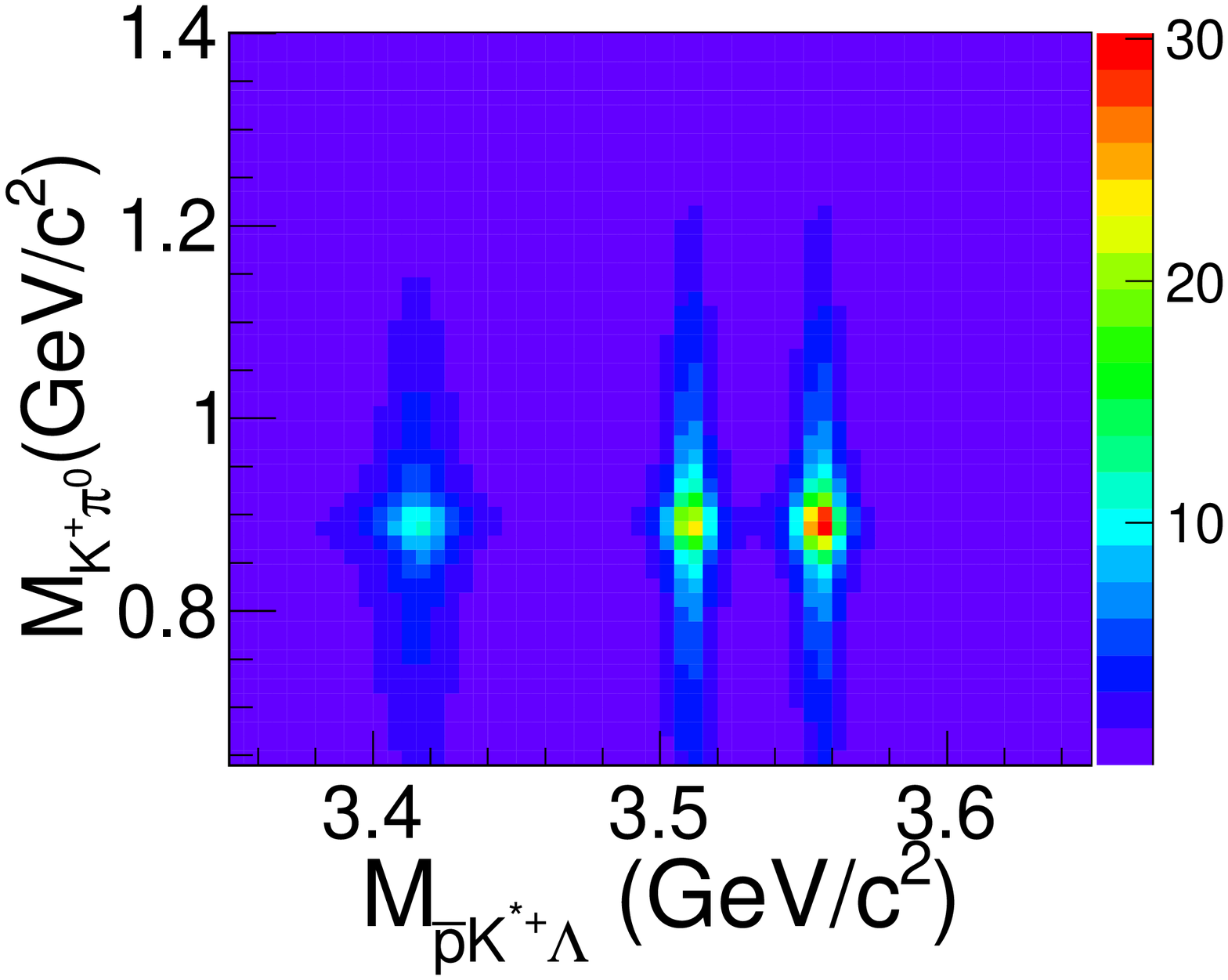}
    \put(-80,70){(b)}
  %\end{minipage}
  }
  \subfigure{
  %\begin{minipage}{0.3\textwidth}
    \centering
    \includegraphics[width=2.2in]{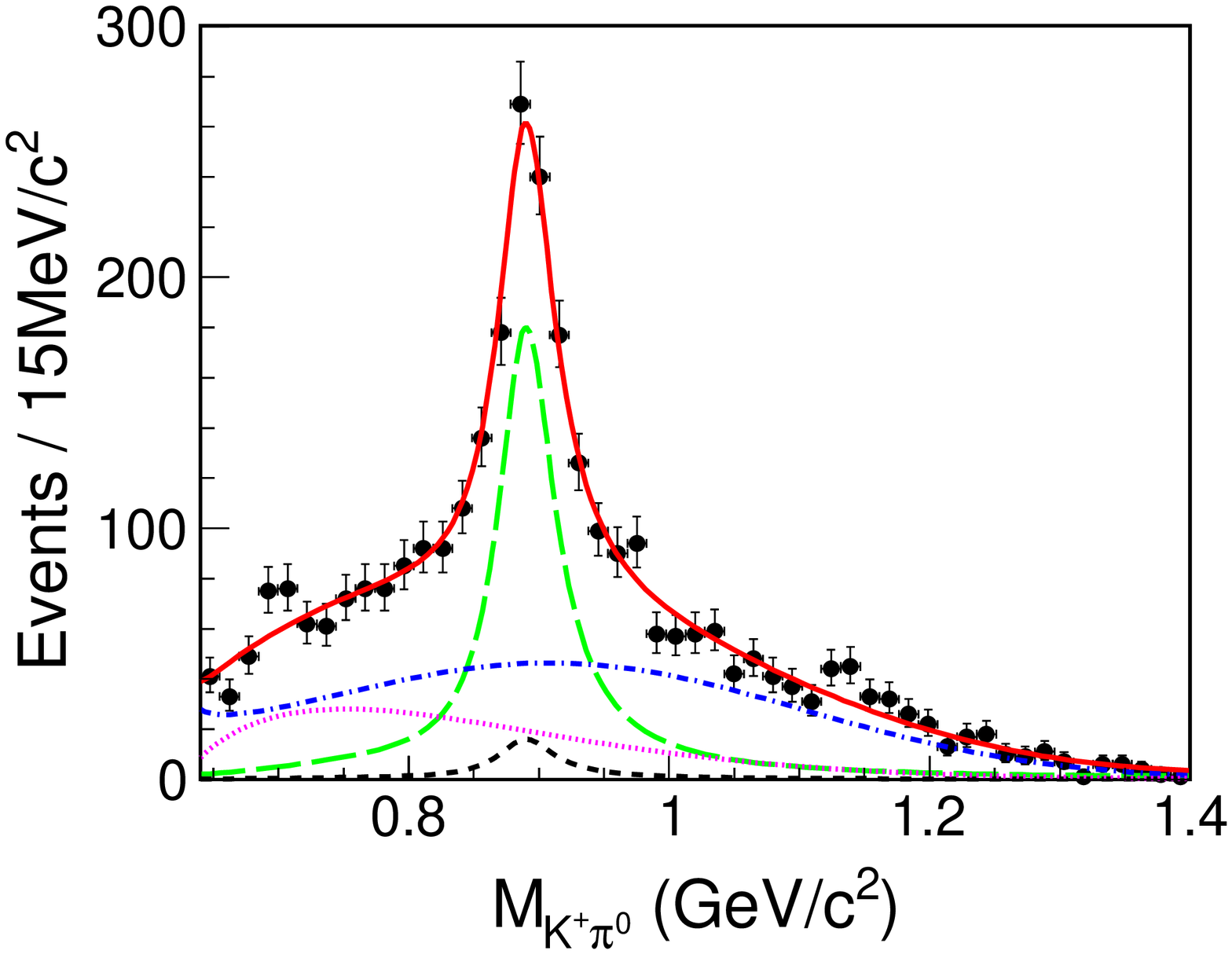}
    \put(-30,80){(c)}
  %\end{minipage}
  }
  \subfigure{
  %\begin{minipage}{0.3\textwidth}
    \centering
    \includegraphics[width=2.2in]{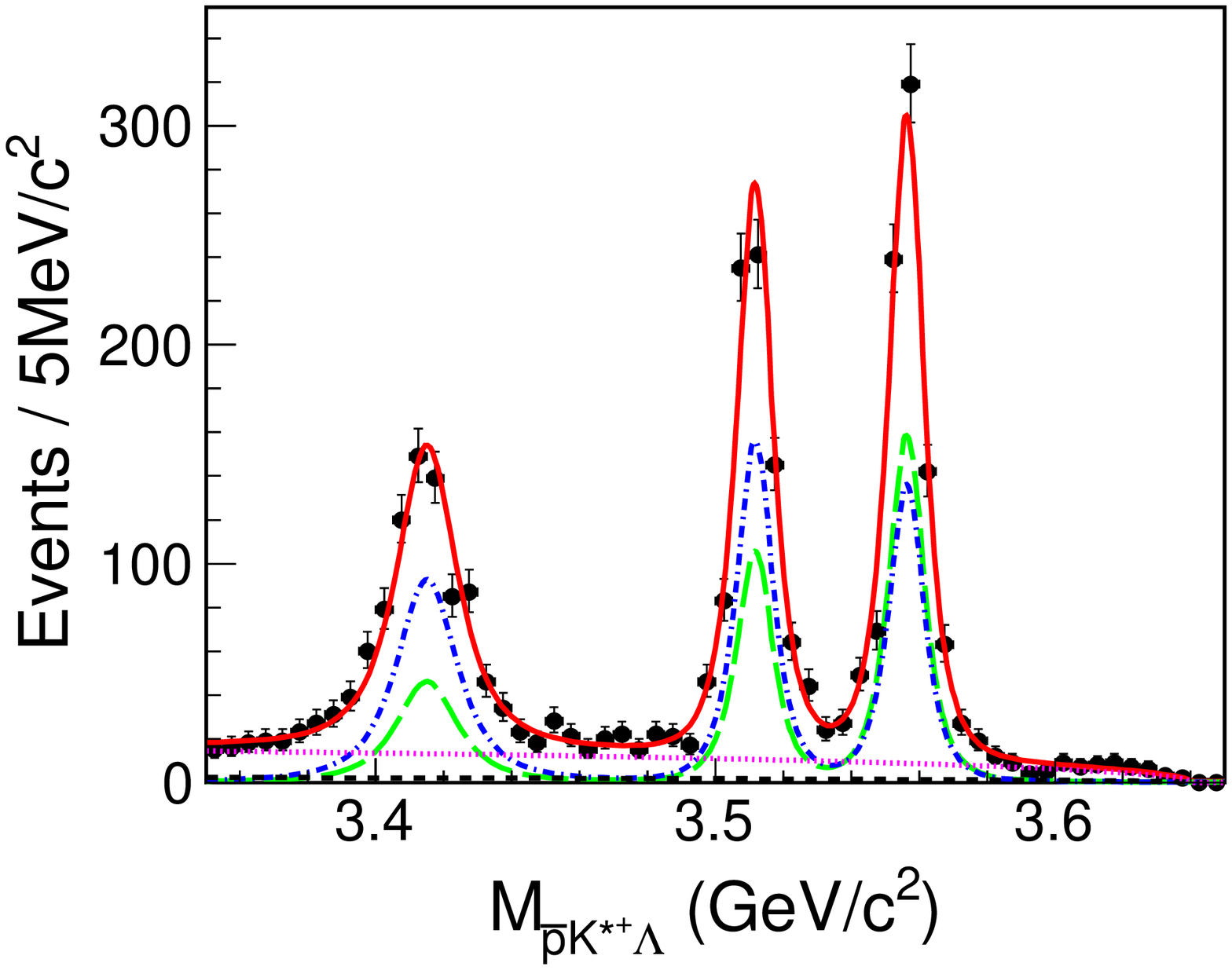}
    \put(-30,80){(d)}
  %\end{minipage}
  }
  \caption{(a) Distribution of $M_{K^{+}\pi^{0}}$ versus $M_{\bar{p}K^{*+}\Lambda}$ from data. The three boxes  indicate from left to right the signal region of $\chi_{c0}$, $\chi_{c1}$, and $\chi_{c2}$, respectively. (b) 2-D histogram sampled from the composite PDF of the  2-D fit. (c) and (d) are projections of the 2-D fit on the distributions of $M_{K^{+}\pi^{0}}$ and $M_{\bar{p}K^{*+}\Lambda}$, respectively. The dots with error bars are data; the solid curves show the fitting result; the long-dashed curves are ($K^{*+}, \chi_{cJ}$) signal; the short-dashed curves are ($K^{*+}$, non-$\chi_{cJ}$) background; the dot-dashed curves are ($\chi_{cJ}$, non-$K^{*+}$) background and the dotted curves are (non-$K^{*+}$, non-$\chi_{cJ}$) background.}\label{fig5}
\end{figure*}

   The distribution of $M_{K^{+}\pi^{0}}$ versus $M_{\bar{p}K^{*+}\Lambda}$ is shown in Fig.\ \ref{fig5}. An unbinned extended maximum-likelihood 2-D fit is performed on this distribution to
determine the number of ($K^{*+}$, $\chi_{cJ}$) events.
   The composite probability density function (PDF) is constructed as follows:
   \begin{equation}
    \begin{split}
      F&=N_{\rm sig}^{\rm obs}\times(F^{K^{*}}_{\rm sig}\cdot F^{\chi_{cJ}}_{\rm sig}) \\
       &+N_{\rm bkg}^{\chi_{cJ},{\rm non}-K^{*}}\times(F^{{\rm non}-K^{*}}_{\rm bkg}\cdot F^{\chi_{cJ}}_{\rm sig})\\
       &+N_{\rm bkg}^{K^{*},{\rm non}-\chi_{cJ}}\times(F^{{\rm non}-\chi_{cJ}}_{\rm bkg}\cdot F^{K^{*}}_{\rm sig})\\
       &+N_{\rm bkg}^{{\rm non}-K^{*}\chi_{cJ}}\times(F^{{\rm non}-K^{*}}_{\rm bkg}\cdot F^{{\rm non}-\chi_{cJ}}_{\rm bkg}).
    \end{split}
   \end{equation}
    Here, $N_{\rm sig}^{\rm obs}$, $N_{\rm bkg}^{\chi_{cJ},{\rm non}-K^{*}}$, $N_{\rm bkg}^{K^{*},{\rm non}-\chi_{cJ}}$, and $N_{\rm bkg}^{{\rm non}-K^{*}\chi_{cJ}}$ are the numbers of ($K^{*}$, $\chi_{cJ}$) signal events, ($\chi_{cJ}$, non-$K^{*}$), ($K^{*}$, non-$\chi_{cJ}$), and (non-$K^{*}$, non-$\chi_{cJ}$) background events, respectively.

The  shape of the $K^{*+}$ resonance, $F^{K^{*}}_{\rm sig}$, is described by a $P$-wave Breit-Wigner (BW) function \cite{P16} convolved with a double-Gaussian function ($DG$) that accounts for detector resolution, the parameters of which are determined from MC simulation. The definition of $F^{K^{*}}_{\rm sig}$ is
   \begin{equation}
     F^{K^{*}}_{\rm sig}(s)=\frac{M\Gamma(s)}{(s^{2}-M^{2})^{2}+M^{2}\Gamma(s)^{2}}\otimes DG(s),
   \end{equation}
   where $\Gamma(s)=\Gamma(\frac{M}{s})^{2}(\frac{q}{q_{0}})^{2L+1}$, $s$ is the invariant mass of the  $K^{+}\pi^{0}$ pair, $M$ and $\Gamma$ are the $K^{*+}$ mass and width~\cite{P10}, $q$ is the $K^{+}$ momentum in the $K^{*+}$ rest frame, $q_{0}$ is the $q$ value at $s=M$, and $L=1$ is the relative orbital angular momentum of $K^{+}\pi^{0}$.

The background distribution of the fake $K^{*+}$ contribution, $F^{{\rm non}-K^{*}}_{\rm bkg}$, is described by truncated polynomial function $F^{{\rm non}-K^{*}}_{\rm bkg}(s)=(s-m_{t})^{a}e^{-bs-cs^{2}}$ \cite{P16}, where $m_{t}$ is the threshold mass for $K^{+}\pi^{0}$ and $a$, $b$, $c$ are free parameters.

The  shape of the $\chi_{cJ}$ signal is described by
   \begin{equation}
     F^{\chi_{cJ}}_{\rm sig}=E_{\gamma}^{3}\cdot f(E_{\gamma})\cdot BW(m)\cdot \frac{B_{l}(Q)}{B_{l}(Q_{0})}\otimes G(m;\mu,\sigma).
   \end{equation}
  Here $E_{\gamma}^{3}$ is an E1 radiative-transition factor and  $f(E_{\gamma})=\frac{E_{0}^{2}}{E_{\gamma}E_{0}+(E_{\gamma}-E_{0})^{2}}$ is a damping factor~\cite{P17}, where $E_{\gamma}$ is the energy of the radiative photon in the $\psi(3686)$ rest frame and $E_{0}=\frac{M_{\psi(3686)}^{2}-M_{\chi_{cJ}}^{2}}{2M_{\psi(3686)}^{2}}$.  In the relativistic BW function  $BW(m)$, the mass and width of the $\chi_{cJ}$ are fixed to the PDG \cite{P10} values.  The Blatt-Weisskopf barrier factor \cite{P18} $B_{l}(Q)$ is a function of $Q$, which is the momentum of either the radiative photon or the $\chi_{cJ}$ in the $\psi(3686)$ rest frame, $Q_{0}$ is the $Q$ value at $m=M_{\chi_{cJ}}$, where $m$ is the invariant mass of the $\bar{p}K^{*+}\Lambda$ combination. Finally, $G(m;\mu,\sigma)$ is a modified Gaussian function parameterizing the instrumental mass resolution, taking the form~\cite{P19}
   \begin{equation}
     G(m;\mu,\sigma)=\frac{1}{\sqrt{2\pi}\sigma} e^{-(|\frac{m-\mu}{\sigma}|)^{1+\frac{1}{1+|\frac{m-\mu}{\sigma}|}}},
   \end{equation}
where the parameters are determined from MC simulation.

   The shape of fake $\chi_{cJ}$ candidates, $F^{{\rm non}-\chi_{cJ}}_{\rm bkg}$, is described by an ARGUS \cite{P20} function.

   The fit yields $254\pm35$
($K^{*+}$, $\chi_{c0}$)
events with a statistical significance of 7.2$\sigma$, $328\pm36$ ($K^{*+}$, $\chi_{c1}$) events with a statistical significance of 11.6$\sigma$, and $476\pm52$ ($K^{*+}$, $\chi_{c2}$) events with a statistical significance of 15.2$\sigma$. The statistical significance is determined from the change of the log-likelihood value and the degrees of freedom in the fit when performed with and without a signal component.
The 2-D histogram sampled from the composite PDF and the projections of the fit on the  $M_{K^{+}\pi^{0}}$ and $M_{\bar{p}K^{*+}\Lambda}$ distributions are shown in Fig.\ \ref{fig5}.

The BF of $\chi_{cJ}\rightarrow\bar{p}K^{*+}\Lambda$ is calculated by
   \begin{equation}\label{2}
    \begin{split}
     \mathcal{B}&=\frac{N_{\rm sig}^{\rm obs}-N_{\rm bkg}}{\epsilon\cdot N_{\psi(3686)}\cdot \mathcal{B}(\psi(3686)\rightarrow\gamma\chi_{cJ})}\\
                &\times\frac{1}{\mathcal{B}(\Lambda\rightarrow p\pi^{-})\cdot\mathcal{B}(K^{*+}\rightarrow K^{+}\pi^{0})\cdot\mathcal{B}(\pi^{0}\rightarrow\gamma\gamma)},
    \end{split}
   \end{equation}
where $N_{\rm sig}^{\rm obs}$ is the number of signal event returned  from the 2-D fit and $N_{\rm bkg}=11.7\pm3.5$, $5.1\pm2.3$, $4.8\pm2.6$ are the numbers of ($K^{*}$, $\chi_{c0}$), ($K^{*}$, $\chi_{c1}$), ($K^{*}$, $\chi_{c2}$) peaking background events, respectively, which is reported in Sec.\ \ref{s3.2}; $N_{\psi(3686)}=(448.1\pm2.9)\times10^{6}$ is the number of $\psi(3686)$ events \cite{P5}, and $\epsilon$ are detection efficiencies which are determined from MC simulation and found to be $(5.51\pm0.05)\%$, $(7.07\pm0.06)\%$, and $(6.31\pm0.06)\%$ for the $\chi_{c0}$, $\chi_{c1}$, and $\chi_{c2}$ signals, respectively. The BFs $\mathcal{B}(\psi(3686)\rightarrow\gamma\chi_{cJ})$ , $\mathcal{B}(\Lambda\rightarrow p\pi^{-})$, $\mathcal{B}(K^{*+}\rightarrow K^{+}\pi^{0})$, and $\mathcal{B}(\pi^{0}\rightarrow\gamma\gamma)$ are taken from Ref.~\cite{P10}. The BFs of $\chi_{cJ}\rightarrow\bar{p}K^{*+}\Lambda$ are measured to be $(4.8\pm0.7)\times10^{-4}$ for the $\chi_{c0}$ mode, $(5.0\pm0.5)\times10^{-4}$ for the $\chi_{c1}$  mode, and $(8.2\pm0.9)\times10^{-4}$ for the $\chi_{c2}$ mode, where the uncertainties are statistical only.

\section{\label{s4}Study of $\bm{\psi(3686)\rightarrow\bar{p}K^{*+}\Lambda}$}
\subsection{\label{s4.1}Event Selection}

Events are selected containing at least two photons, one $\bar{p}$, one $K^+$, and one $\Lambda$ candidate, identified using the same criteria as employed in  the $\psi(3686)\rightarrow\gamma\bar{p}K^{*+}\Lambda$ analysis. The selected particles are subjected to a 5C kinematic fit under the hypothesis of $\psi(3686)\rightarrow\bar{p}K^{+}\pi^{0}\Lambda$, with the invariant mass of the two photons constrained to the $\pi^{0}$ mass. The $\chi^{2}$ of the 5C fit is required to be less than 100. For events with more than one combination meeting this requirement, only the combination with the smallest $\chi^{2}$ is retained for further analysis. To veto backgrounds from $\psi(3686)\rightarrow\gamma\bar{p}K^{+}\pi^{0}\Lambda$ and $\psi(3686)\rightarrow\gamma\bar{p}K^{+}\Lambda$, an alternative 5C (4C) kinematic fit is performed under the $\psi(3686)\rightarrow\gamma\bar{p}K^{+}\pi^{0}\Lambda$ ($\gamma\bar{p}K^{+}\Lambda$) hypothesis. We further require that the confidence level of the kinematic fit for the $\psi(3686)\rightarrow\bar{p}K^{+}\pi^{0}\Lambda$ assignment is larger than those of  the $\psi(3686)\rightarrow\gamma\bar{p}K^{+}\pi^{0}\Lambda$ and $\psi(3686)\rightarrow\gamma\bar{p}K^{+}\Lambda$ hypotheses. \par

\begin{figure}
  \centering
  \subfigure{
  %\begin{minipage}{0.22\textwidth}
    \centering
    \includegraphics[width=1.5in]{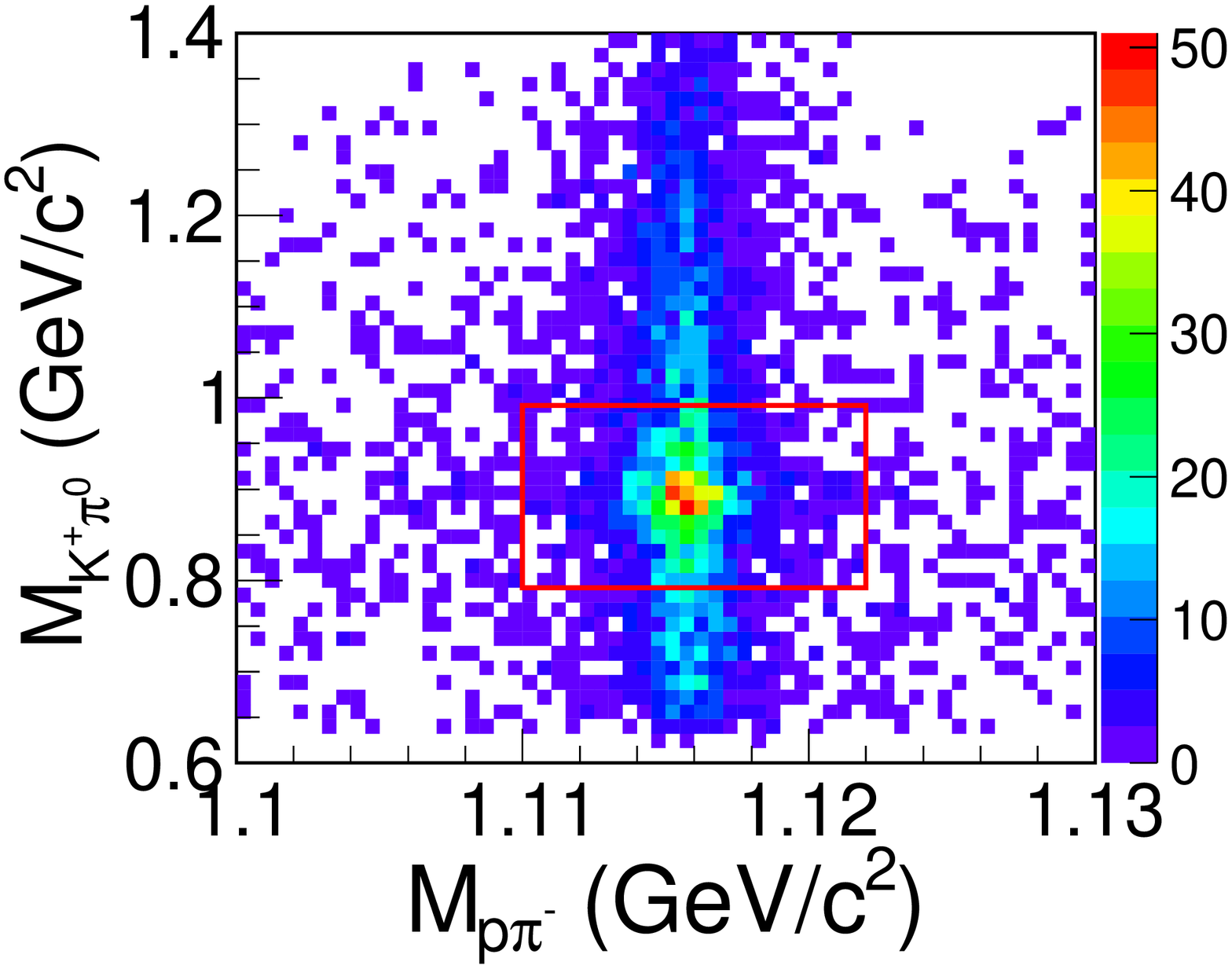}
    \put(-28,73){(a)}
  %\end{minipage}
  }\hspace{-0.05in}
  \subfigure{
  %\begin{minipage}{0.22\textwidth}
    \centering
    \includegraphics[width=1.5in]{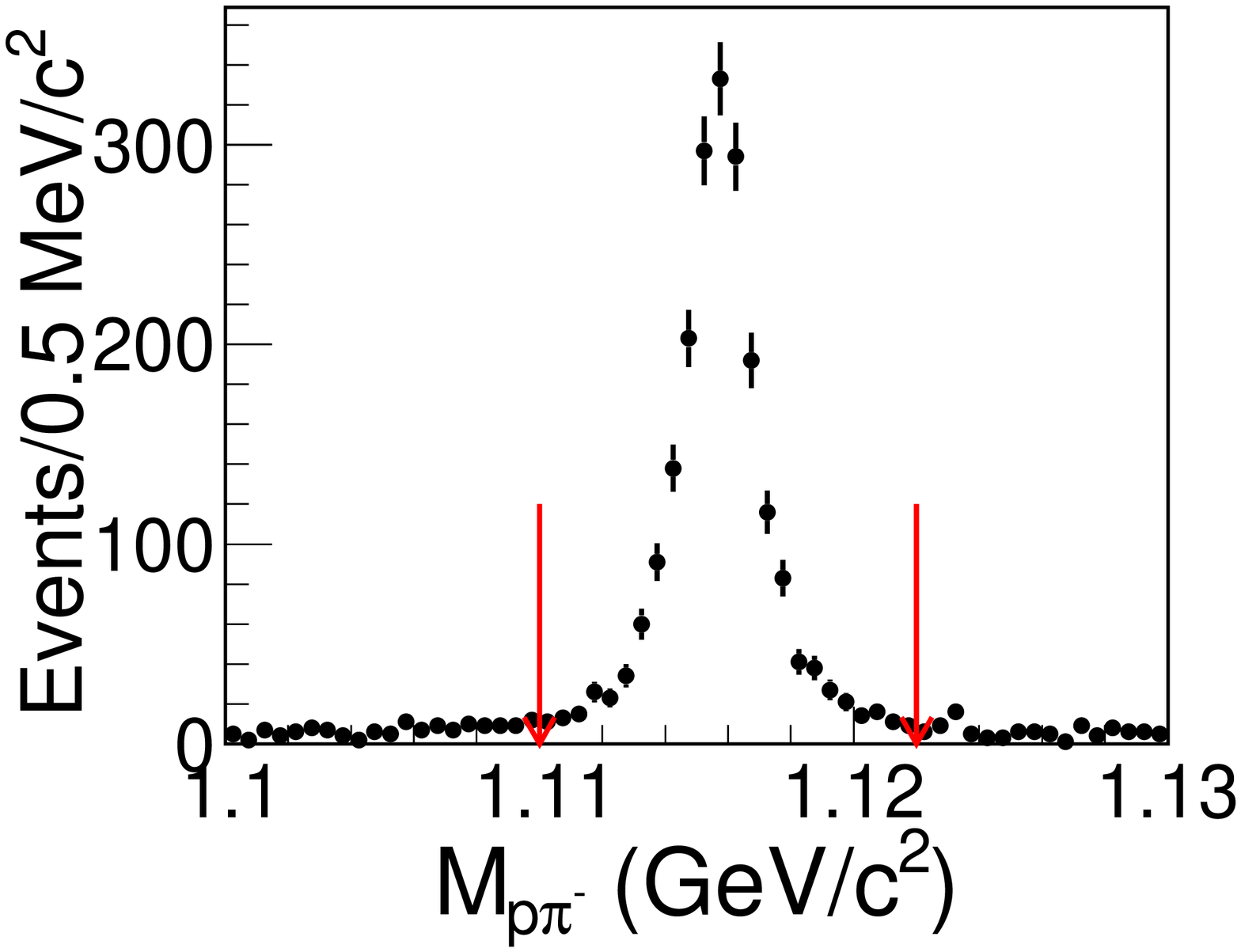}
    \put(-28,73){(b)}
  %\end{minipage}
  }
  \caption{(a) Distribution of $M_{K^{+}\pi^{0}}$ versus $M_{p\pi^{-}}$. The box indicates the signal region. (b) Invariant mass distribution of $p\pi^{-}$. The arrows indicates the mass window used in the selection.}\label{fig9}
\end{figure}

\begin{figure*}{}
  \centering
  \subfigure{
  %\begin{minipage}{0.33\textwidth}
    \centering
    \includegraphics[width=2.2in]{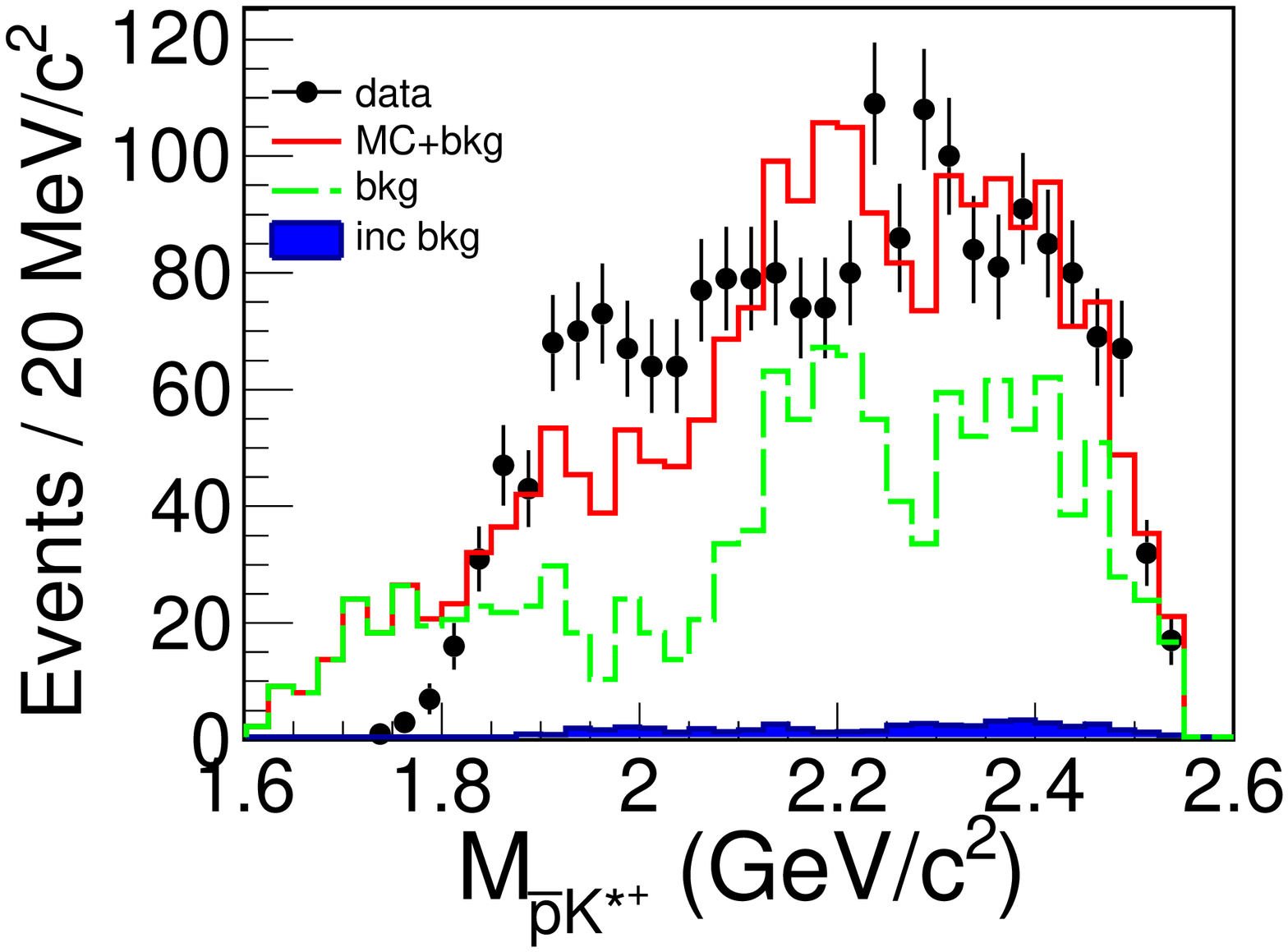}
    \put(-25,90){(a)}
  %\end{minipage}
  }
  \subfigure{
  %\begin{minipage}{0.33\textwidth}
    \centering
    \includegraphics[width=2.2in]{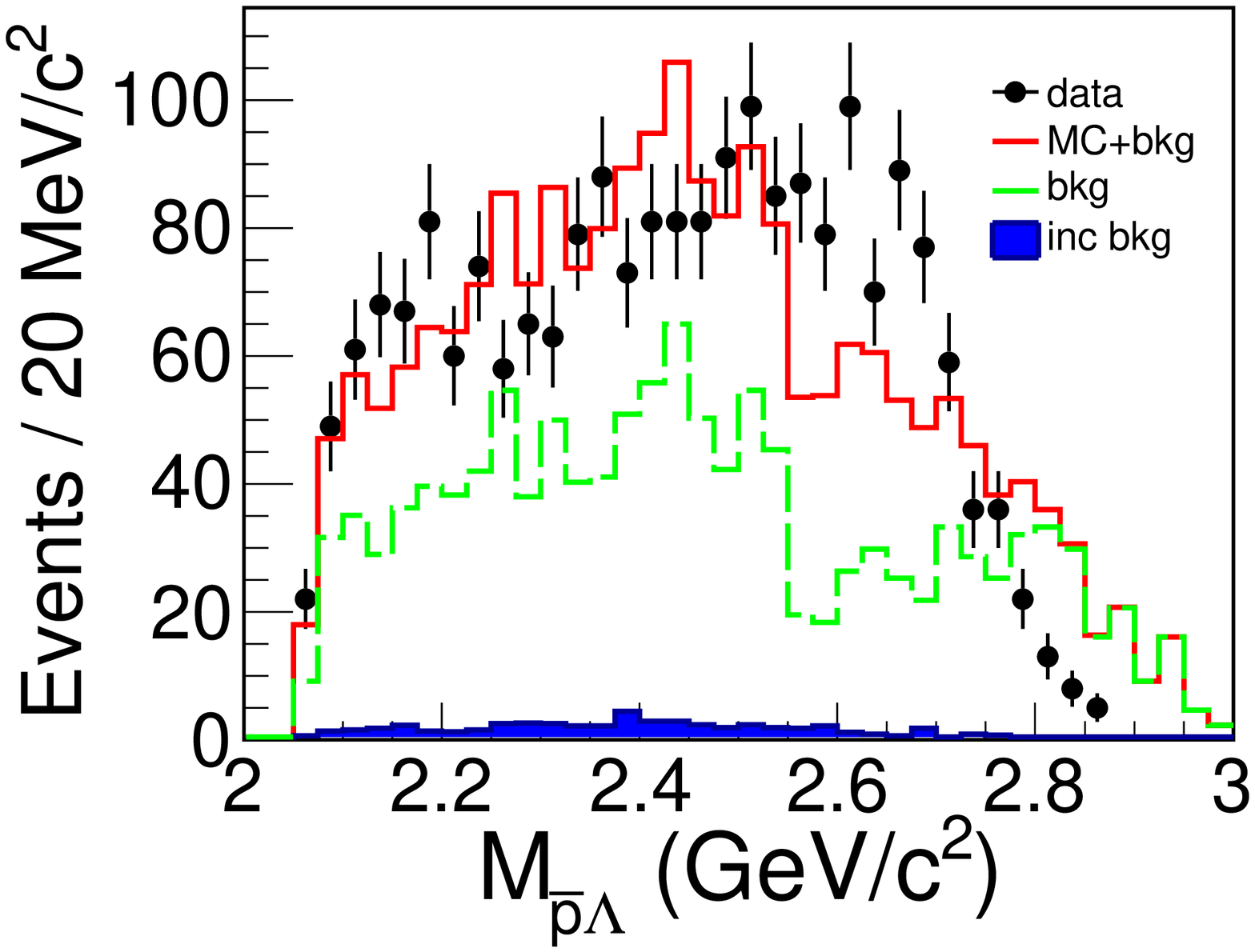}
    \put(-110,90){(b)}
  %\end{minipage}
  }
  \subfigure{
  %\begin{minipage}{0.33\textwidth}
    \centering
    \includegraphics[width=2.2in]{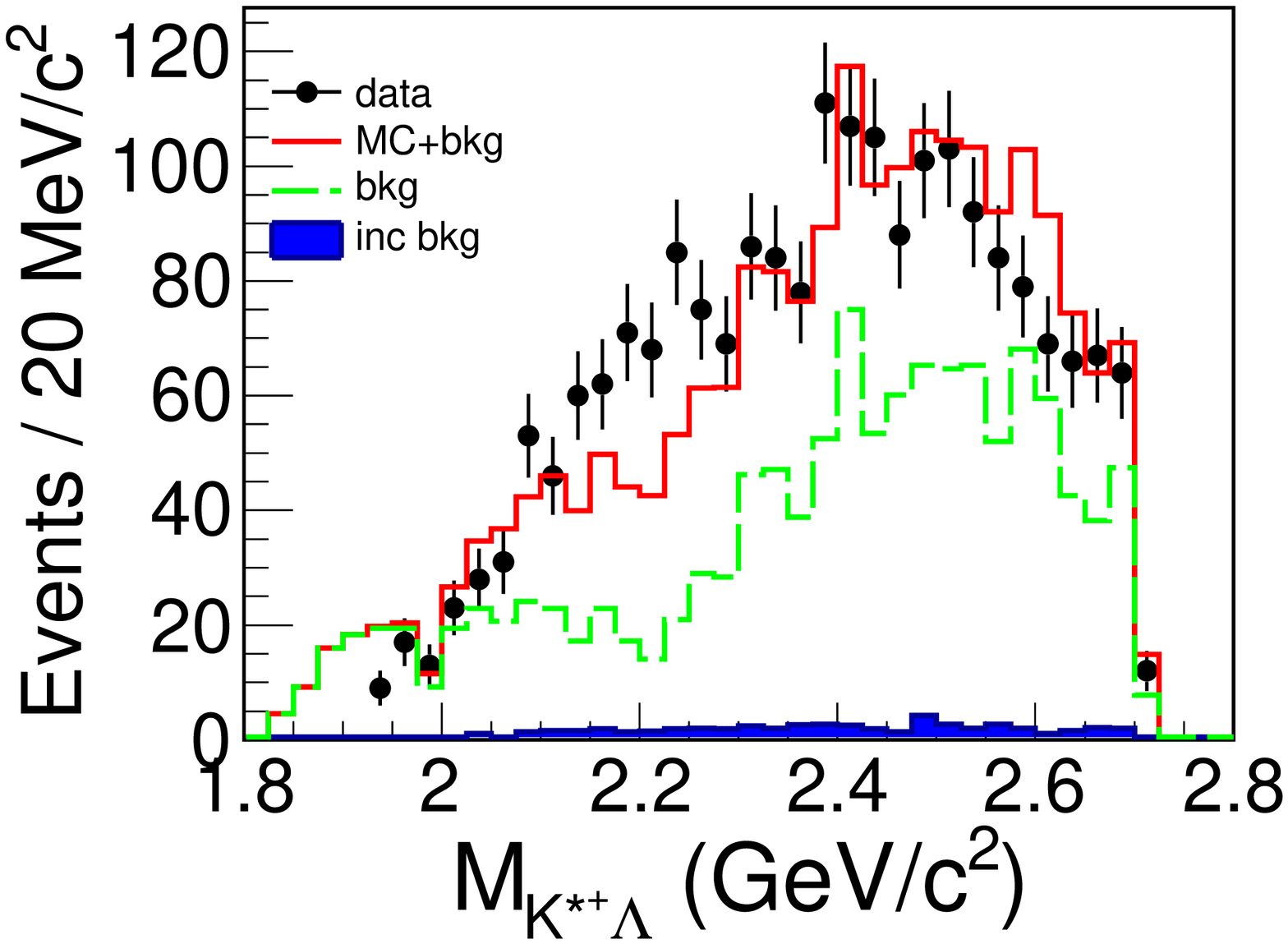}
    \put(-25,90){(c)}
  %\end{minipage}
  }
  \subfigure{
  %\begin{minipage}{0.32\textwidth}
  \centering
  \includegraphics[width=2.2in]{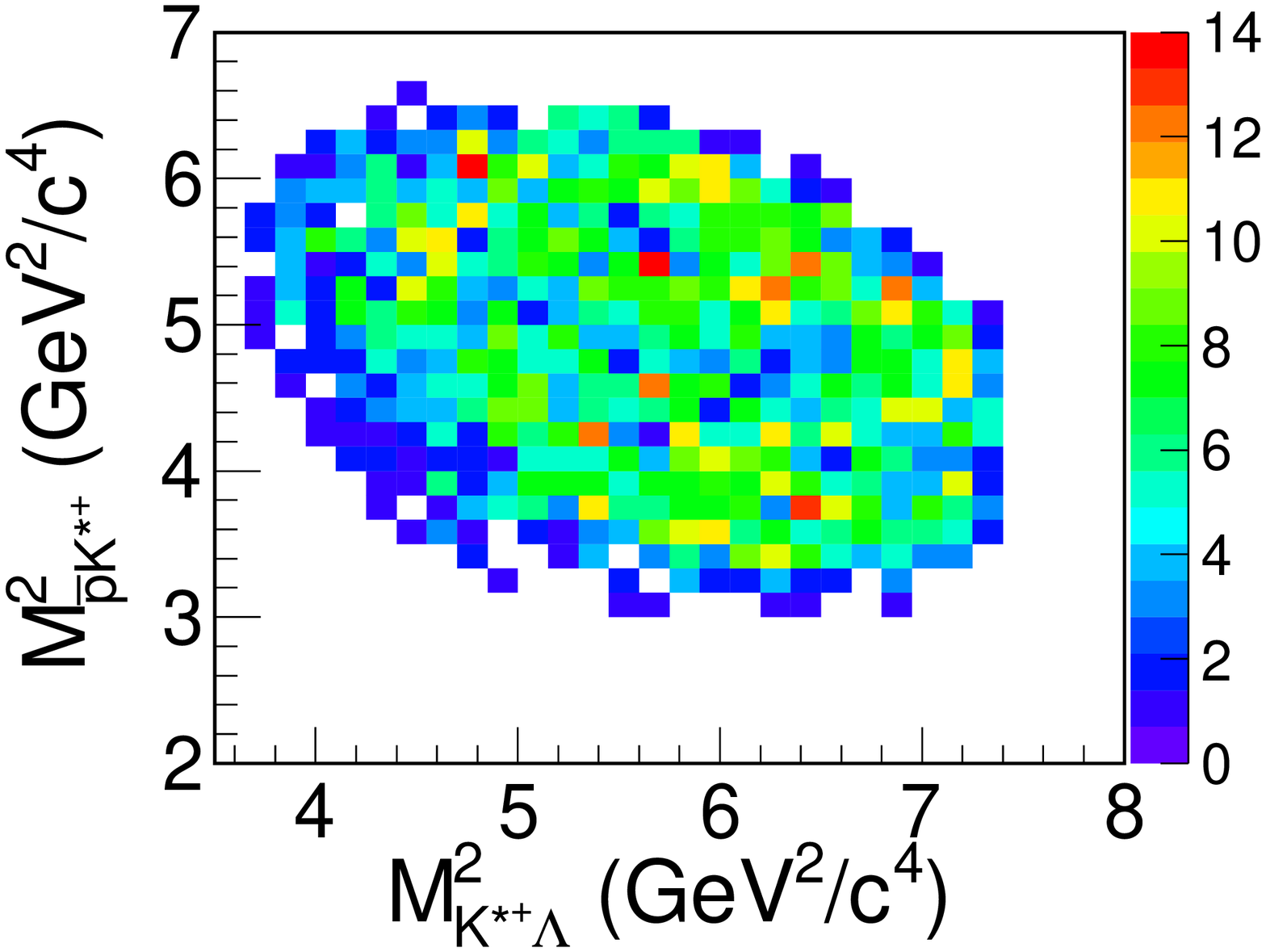}
  \put(-35,90){(d)}
  %\end{minipage}
  }
  \caption{Invariant mass spectra of (a) $M_{\bar{p}K^{*+}}$, (b) $M_{\bar{p}\Lambda}$, and (c) $M_{K^{*+}\Lambda}$. The dots with error bars are data. The shaded histograms are background from inclusive MC sample. The dashed lines are background that are estimated from the $K^{*+}$ sidebands and are normalized to the signal region. The solid lines are the sum of phase-space MC sample and non-$K^{*+}$ background that are normalized to signal yields. (d) Dalitz plot of $\bar{p}K^{*+}\Lambda$.}\label{fig11}
\end{figure*}

   The distribution of $M_{K^{+}\pi^{0}}$ versus $M_{p\pi^{-}}$ is shown in Fig.\ \ref{fig9}(a), where $K^{*+}$ and $\Lambda$ signals are visible. The $\Lambda$ candidates are selected by requiring $|M_{p\pi^{-}}-M_{\Lambda}|<6~{\rm MeV}/c^{2}$ and $K^{*+}$ candidates are selected by requiring $|M_{K^{+}\pi^{0}}-M_{K^{*+}}|<0.1~{\rm GeV}/c^{2}$. The $K^{*+}$ sidebands are defined to be $1.1<M_{K^{+}\pi^{0}}<1.2~{\rm GeV}/c^{2}$ and $0.65<M_{K^{+}\pi^{0}}<0.75~{\rm GeV}/c^{2}$. The distribution of $M_{p\pi^{-}}$ for events within the $K^{*+}$ signal region is shown in Fig.\ \ref{fig9}(b). The mass spectra of $\bar{p}K^{*+}$, $\bar{p}\Lambda$, $K^{*+}\Lambda $, and Dalitz plot after the application of all selection criteria are shown in Fig.\ \ref{fig11}.
A near-threshold structure in the $M_{\bar{p}\Lambda}$ is fitted with a 1.7$\sigma$ signficance, using the the same parameterization as in the $\chi_{cJ}\rightarrow\bar{p}K^{*+}\Lambda$ analysis.\par

\subsection{\label{s4.2}Background study}

 Using an inclusive MC sample of $506\times10^{6}$ $\psi(3686)$ events, the background from fake $\Lambda$ is found together with fake $K^{*+}$. The sources of background can be categorized into two types: peaking background events with genuine $K^{*+}$ mesons in the final state and non-peaking background events with fake $K^{*+}$ candidates. The non-peaking background can be estimated from a fit to the $M_{K^{+}\pi^{0}}$ spectrum. The major peaking backgrounds are found to be: $\psi(3686)\rightarrow\gamma\chi_{cJ}\rightarrow\gamma\bar{p}K^{*+}\Lambda$ ($J$=0, 1, 2) and $\psi(3686)\rightarrow\bar{p}K^{*+}\Sigma^{0},\Sigma^{0}\rightarrow\gamma\Lambda$. Corresponding exclusive MC samples are generated for further studies. The selection criteria are applied to these exclusive MC samples and the number of surviving events are normalized by the BFs
  of the relevant decay processes.
% For the normalization procedure, the $\mathcal{B}%(\psi(3686)\rightarrow\gamma\chi_{cJ}\rightarrow\gamma\bar{p}K^{*+}\Lambda)$ have been measured in Sec.\ \ref{s3.3} and {\color{red} $\mathcal{B}(\psi(3686)\rightarrow\bar{p}K^{*+}\Sigma^{0})$ is roughly measured to be $(0.6\pm0.2)\times10^{-6}$}.
The normalized number of $\psi(3686)\rightarrow\bar{p}K^{*+}\Sigma^{0}$ background events is 5.2$\pm$1.1 and the expected numbers of $\psi(3686)\rightarrow\gamma\chi_{cJ}\rightarrow\gamma\bar{p}K^{*+}\Lambda$ ($J$=0, 1, 2) background decays are 1.9$\pm$0.3, 4.5$\pm$0.5 and 8.8$\pm$1.0, respectively.
%These peaking background events will be subtracted.\par

 A data sample of 2.93 fb$^{-1}$ \cite{P15} collected at $\sqrt{s}=3.77~{\rm GeV}$ is used to investigate possible background from continuum processes. After normalizing to the integrated luminosity of the $\psi(3686)$ data sample, 164.1$\pm$9.5 events survive and a clear $K^{*+}$ peak is found in the  $K^{+}\pi^{0}$ mass spectrum.
 This background yield is cross-checked by repeating the procedure on the data sample of  44.5 pb$^{-1}$ \cite{P21} collected at $\sqrt{s}=3.65$ GeV, and a compatible result of  207$\pm$61 events is obtained, after normalization.

\subsection{\label{s4.3}Branching fraction measurement of $\bm{\psi(3686)\rightarrow\bar{p}K^{*+}\Lambda}$}

\begin{figure}
  \centering
  \includegraphics[width=0.3\textwidth]{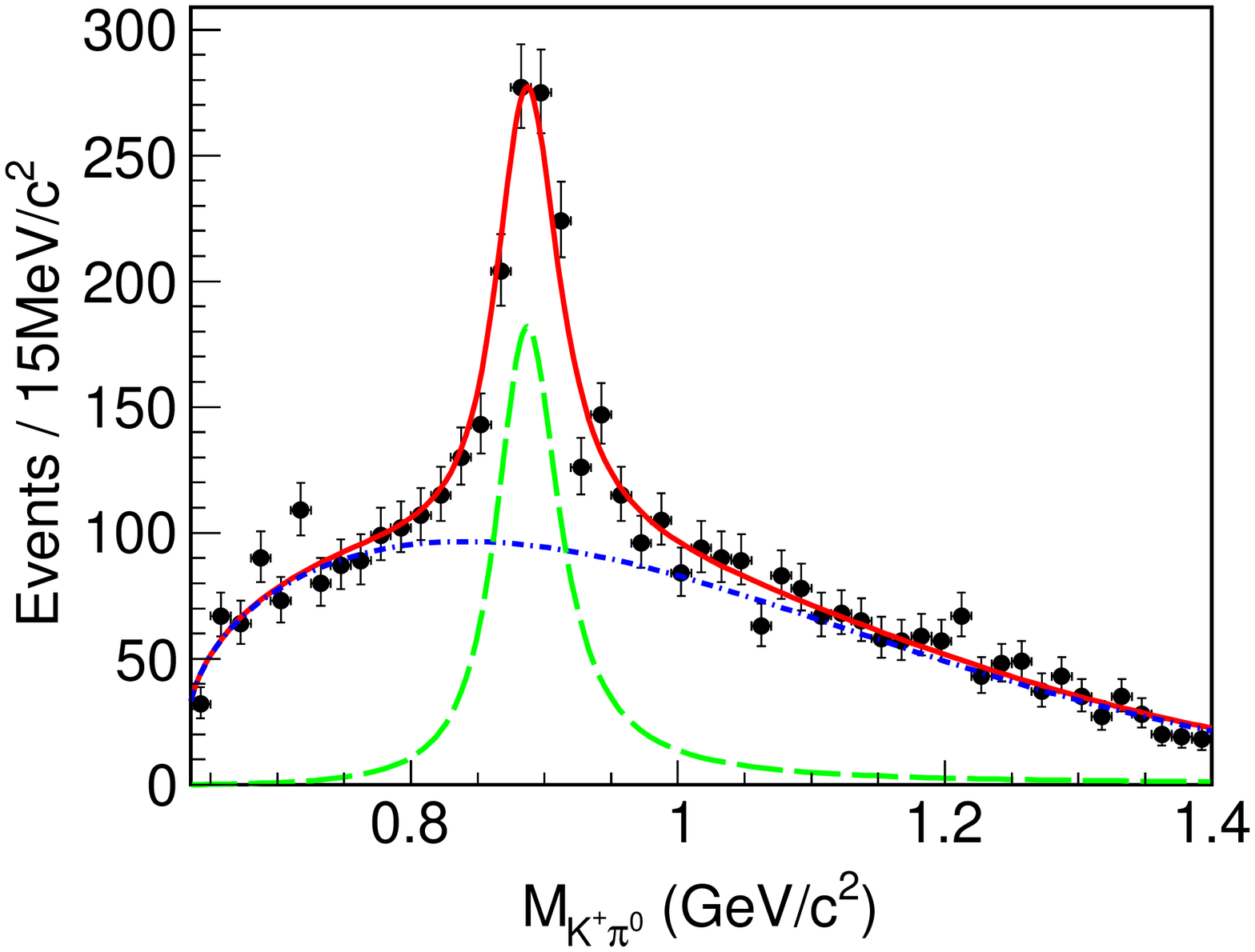}\\
  \caption{Invariant-mass spectrum of $K^{+}\pi^{0}$, showing the fit result. The dots with error bars are data and the solid curve shows the fit. The short-dashed curve is $K^{*+}$ signal and the long-dashed curve is non-peaking background.}\label{fig15}
\end{figure}

   An unbinned maximum likelihood fit is performed to the distribution of $M_{K^{+}\pi^{0}}$ (Fig.\ \ref{fig15}) to extract the number of $K^{*+}$ signal events. The $K^{*+}$ signal shape is described by a $P$-wave BW function convolved with a double-Gaussian function, and the background shape is described by a truncated polynomial function. The definitions of these functions are the same as those introduced in Sec.\ \ref{s3.3}. The fit result is shown in  Fig.\ \ref{fig15}.

The BF of $\psi(3686)\rightarrow\bar{p}K^{*+}\Lambda$ is calculated according to
   \begin{equation}\label{2}
    \begin{split}
     \mathcal{B}&=\frac{N_{\rm sig}^{\rm obs}-N_{\rm bkg}}{\epsilon\cdot N_{\psi(3686)}\cdot \mathcal{B}(\Lambda\rightarrow p\pi^{-})}\\
                &\times\frac{1}{\mathcal{B}(K^{*+}\rightarrow K^{+}\pi^{0})\cdot\mathcal{B}(\pi^{0}\rightarrow\gamma\gamma)},
    \end{split}
   \end{equation}
   where $N_{\rm sig}^{\rm obs}= 1011\pm60 $ is number of $K^{*+}$ signal events obtained from the fit, $N_{\rm bkg}=20.4\pm1.6$ is the number of peaking background events reported in Sec.\ \ref{s4.2}, and $\epsilon$ is the detection efficiency, $(14.0\pm0.1)\%$,  estimated from MC simulation. The $\mathcal{B}(\psi(3686)\rightarrow\bar{p}K^{*+}\Lambda)$ is measured to be $(6.3\pm0.5)\times10^{-5}$, where the uncertainty is statistical only.

\section{\label{s5}Systematic uncertainties}

\begin{table*}[t]
  \centering
  \caption{Summary of systematic uncertainties (in \%) in the measured BFs of $\chi_{cJ}\rightarrow\bar{p}K^{*+}\Lambda$ and $\psi(3686)\rightarrow\bar{p}K^{*+}\Lambda$.}\label{table3}
  \begin{tabular}{p{4.cm}p{2.cm}p{2.cm}p{2.cm}c}
  \hline
  \hline
  Source &\multicolumn{3}{c}{$\chi_{cJ}\rightarrow\bar{p}K^{*+}\Lambda$} &$\psi(3686)\rightarrow\bar{p}K^{*+}\Lambda$\\ \hline
  &$\chi_{c0}$ &$\chi_{c1}$ &$\chi_{c2}$\\ \hline
  MDC Tracking &4.0 &4.0 &4.0 &4.0\\
  PID efficiency &4.0 &4.0 &4.0 &4.0\\
  Photon detection &3.0 &3.0 &3.0 &2.0\\
  $\Lambda$ mass window &0.1 &0.1 &0.1 &0.1\\
  Kinematic fit &0.1 &0.5 &0.2 &1.4\\
  Fit range &5.9 &2.1 &2.0 &3.0\\
  Signal shape &4.9 &3.8 &4.1 &3.4\\
  Background shape &1.3 &2.0 &0.7 &1.1\\
  Number of $\psi(3686)$ events &0.7 &0.7 &0.7 &0.7\\
  $\mathcal{B}(\Lambda\rightarrow p\pi^{-})$ &0.8 &0.8 &0.8 &0.8\\
  $\mathcal{B}(\psi(3686)\rightarrow\gamma\chi_{cJ})$ &2.0 &2.5 &2.1 &--\\  \hline
  Total &10.3 &8.5 &8.2 &7.8\\ \hline
  \hline
  \end{tabular}
\end{table*}
%   The systematic uncertainties in the BF measurements originate mainly from knowledge of the  tracking efficiency, PID, photon reconstruction,  the $\Lambda$ mass window requirement, the kinematic fit, the fit range, the choice of signal and background shapes,  the number of $\psi( 3686)$ events, and the BFs of the intermediate state decays.
Systematic uncertainties on the BF measurements arise from a variety of sources:

   \textit{Tracking efficiency}. The uncertainty due to data-MC difference in the tracking efficiency is 1\% for each charged track coming from a primary vertex according to a study of $J/\psi\rightarrow K^{*}\bar{K}$ and $J/\psi\rightarrow p\bar{p}\pi^{+}\pi^{-}$ events. For each track from $\Lambda$, the uncertainty is also 1\% from analysis of $J/\psi\rightarrow \bar{p}K^{+}\Lambda$ events~\cite{P4}.\par
   \textit{PID efficiency}. The candidates require tracks to be identified as $p$, $\bar{p}$, $K^{+}$, or $\pi^{-}$. The PID efficiency have been investigated using control samples of $J/\psi\rightarrow K^{0}_{S}K^{\pm}\pi^{\pm}$ and $J/\psi\rightarrow p\bar{p}\pi^{+}\pi^{-}$~\cite{P22,P23}. The uncertainty is assigned to be 1\% per charged track.\par
   \textit{Photon detection efficiency}. The photon detection efficiency was studied in the analysis of $J/\psi\rightarrow\rho\pi$ decays~\cite{P22}. The difference in the detection efficiency between the data and MC simulation is taken as the systematic uncertainty from this source, and $1\%$ is assigned for each photon.\par
   \textit{$\Lambda$ Mass window.} The systematic uncertainty from the requirement on the $\Lambda$ signal region is estimated by smearing the  $p\pi^{-}$  invariant mass in the signal MC sample with a Gaussian function to compensate for the resolution difference between data and MC simulation. The smearing parameters are determined by fitting the $\Lambda$ distribution in data with the MC shape convolved with a Gaussian function. The difference in the detection efficiency as determined from signal MC sample with and without the extra smearing is taken as the systematic uncertainty. \par
   \textit{Kinematic fit.} The systematic uncertainty due to kinematic fitting is estimated by correcting the helix parameters of charged tracks according the method described in Ref.\ \cite{P24}. The differences in the detection efficiency between the MC samples  with and without this correction are taken as the uncertainties, which are $0.1\%$, $0.5\%$, and $0.2\%$ for $\chi_{cJ}\rightarrow\bar{p}K^{*+}\Lambda$ ($J$=0, 1, 2) and $1.4\%$ for $\psi(3686)\rightarrow\bar{p}K^{*+}\Lambda$.\par
   \textit{Fit range}. To estimate the systematic uncertainty due to fit range, several alternative fits in different ranges are performed. The resulting largest difference in the BF is assigned as the systematic uncertainty.\par
   \textit{Signal shape}. To estimate the uncertainty due to the choice of signal shape, the $K^{*+}$ and $\chi_{cJ}$ signal line shapes are replaced by alternative fits using MC shapes and the resulting differences in the BFs are assigned as systematic uncertainties.\par
   \textit{Background shape}. In the measurements of $\mathcal{B}(\chi_{cJ}\rightarrow\bar{p}K^{*+}\Lambda)$ and $\mathcal{B}(\psi(3686)\rightarrow\bar{p}K^{*+}\Lambda)$, the $\chi_{cJ}$ background shape is described by an ARGUS function and the $K^{*+}$ background shape is described by a second-order truncated polynomial function. To estimate the systematic uncertainty due to choice of background shape, an alternative fit is performed in which the ARGUS function is replaced with a second-order Chebychev polynomial function and the $K^{*+}$ signal is described with a third-order truncated polynomial. The change in the measured BF is assigned as the corresponding systematic uncertainty.\par
   \textit{Others.} The uncertainty due to the number of $\psi(3686)$ events is $0.7\%$ \cite{P5}. The systematic uncertainties associated with the intermediate-decay BFs of $\psi(3686)\rightarrow\gamma\chi_{cJ}$ and $\Lambda \rightarrow p\pi^{-}$ are taken from the PDG~\cite{P10}.\par
   The above systematic uncertainties are summarized in Table\ \ref{table3}. The total systematic uncertainty is calculated by assuming the individual components to be independent, and adding their magnitude in quadrature.\par

\section{\label{s6}Results And Summary}
\begin{table}[H]
  \centering
   \caption{The BFs of $\psi(3686)\rightarrow\gamma\chi_{cJ}\rightarrow\gamma\bar{p}K^{*+}\Lambda$, $\chi_{cJ}\rightarrow\bar{p}K^{*+}\Lambda$, and $\psi(3686)\rightarrow\bar{p}K^{*+}\Lambda$, where the first uncertainties are statistical and the second ones systematic.}\label{table5}
  \begin{tabular}{lcc}
  \hline
  \hline
  Decay channel & &Branching fraction\\ \hline
  $\psi(3686)\rightarrow\gamma\chi_{c0}\rightarrow\gamma\bar{p}K^{*+}\Lambda$ & &$(4.7\pm0.7\pm0.5)\times10^{-5}$ \\ \hline
  $\psi(3686)\rightarrow\gamma\chi_{c1}\rightarrow\gamma\bar{p}K^{*+}\Lambda$ & &$(4.8\pm0.5\pm0.4)\times10^{-5}$ \\ \hline
  $\psi(3686)\rightarrow\gamma\chi_{c2}\rightarrow\gamma\bar{p}K^{*+}\Lambda$ & &$(7.8\pm0.9\pm0.6)\times10^{-5}$ \\ \hline
  $\chi_{c0}\rightarrow\bar{p}K^{*+}\Lambda$ & &$(4.8\pm0.7\pm0.5)\times10^{-4}$ \\ \hline
  $\chi_{c1}\rightarrow\bar{p}K^{*+}\Lambda$ & &$(5.0\pm0.5\pm0.4)\times10^{-4}$ \\ \hline
  $\chi_{c2}\rightarrow\bar{p}K^{*+}\Lambda$ & &$(8.2\pm0.9\pm0.7)\times10^{-4}$ \\ \hline
  $\psi(3686)\rightarrow\bar{p}K^{*+}\Lambda$ & &$(6.3\pm0.5\pm0.5)\times10^{-5}$\\ \hline
  \hline
  \end{tabular}
\end{table}

The processes $\psi(3686)\rightarrow\gamma\chi_{cJ}\rightarrow\gamma\bar{p}K^{*+}\Lambda$ and $\psi(3686)\rightarrow\bar{p}K^{*+}\Lambda$ are observed for the first time, using $448.1\times10^{6}$ $\psi(3686)$ events collected with the BESIII detector.  Measurements of the $\mathcal{B}(\psi(3686)\rightarrow\gamma\chi_{cJ})\cdot\mathcal{B}(\chi_{cJ}\rightarrow\bar{p}K^{*+}\Lambda)$ and $\mathcal{B}(\psi(3686)\rightarrow\bar{p}K^{*+}\Lambda)$ are performed, for which the results are listed in Table\ \ref{table5}. For the processes of $\chi_{cJ}\rightarrow\bar{p}K^{*+}\Lambda$ ($J$=0, 1, 2) and $\psi(3686)\rightarrow\bar{p}K^{*+}\Lambda$,
no significant substructure is observed in the invariant-mass spectra of $\bar{p}K^{*+}$ and $K^{*+}\Lambda$.
The  $\bar{p}\Lambda$ mass spectrum is also compatible with the absence of substructure, although fits for possible excesses in the threshold region return results of around two sigma significance in each of the four cases.
The new measurements provide more information for understanding the mechanisms of charmonium decays.

   \par

\begin{acknowledgements}
The BESIII collaboration thanks the staff of BEPCII and the IHEP computing center for their strong support. This work is supported in part by National Key Basic Research Program of China under Contract No. 2015CB856700; National Natural Science Foundation of China (NSFC) under Contracts Nos. 11335008, 11425524, 11625523, 11635010, 11735014; the Chinese Academy of Sciences (CAS) Large-Scale Scientific Facility Program; the CAS Center for Excellence in Particle Physics (CCEPP); Joint Large-Scale Scientific Facility Funds of the NSFC and CAS under Contracts Nos. U1532257, U1532258, U1732263; CAS Key Research Program of Frontier Sciences under Contracts Nos. QYZDJ-SSW-SLH003, QYZDJ-SSW-SLH040; 100 Talents Program of CAS; INPAC and Shanghai Key Laboratory for Particle Physics and Cosmology; German Research Foundation DFG under Contract No. Collaborative Research Center CRC 1044, FOR 2359; Istituto Nazionale di Fisica Nucleare, Italy; Koninklijke Nederlandse Akademie van Wetenschappen (KNAW) under Contract No. 530-4CDP03; Ministry of Development of Turkey under Contract No. DPT2006K-120470; National Science and Technology fund; The Knut and Alice Wallenberg Foundation (Sweden) under Contract No. 2016.0157; The Royal Society, UK under Contract No. DH160214; The Swedish Research Council; U. S. Department of Energy under Contracts Nos. DE-FG02-05ER41374, DE-SC-0010118, DE-SC-0012069; University of Groningen (RuG) and the Helmholtzzentrum fuer Schwerionenforschung GmbH (GSI), Darmstadt
\end{acknowledgements}
\bibliography{bibfile}

\providecommand{\noopsort}[1]{}\providecommand{\singleletter}[1]{#1}%
\begin{thebibliography}{23}%
\makeatletter
\providecommand \@ifxundefined [1]{%
 \@ifx{#1\undefined}
}%
\providecommand \@ifnum [1]{%
 \ifnum #1\expandafter \@firstoftwo
 \else \expandafter \@secondoftwo
 \fi
}%
\providecommand \@ifx [1]{%
 \ifx #1\expandafter \@firstoftwo
 \else \expandafter \@secondoftwo
 \fi
}%
\providecommand \natexlab [1]{#1}%
\providecommand \enquote  [1]{``#1''}%
\providecommand \bibnamefont  [1]{#1}%
\providecommand \bibfnamefont [1]{#1}%
\providecommand \citenamefont [1]{#1}%
\providecommand \href@noop [0]{\@secondoftwo}%
\providecommand \href [0]{\begingroup \@sanitize@url \@href}%
\providecommand \@href[1]{\@@startlink{#1}\@@href}%
\providecommand \@@href[1]{\endgroup#1\@@endlink}%
\providecommand \@sanitize@url [0]{\catcode `\\12\catcode `\$12\catcode
  `\&12\catcode `\#12\catcode `\^12\catcode `\_12\catcode `\%12\relax}%
\providecommand \@@startlink[1]{}%
\providecommand \@@endlink[0]{}%
\providecommand \url  [0]{\begingroup\@sanitize@url \@url }%
\providecommand \@url [1]{\endgroup\@href {#1}{\urlprefix }}%
\providecommand \urlprefix  [0]{URL }%
\providecommand \Eprint [0]{\href }%
\providecommand \doibase [0]{http://dx.doi.org/}%
\providecommand \selectlanguage [0]{\@gobble}%
\providecommand \bibinfo  [0]{\@secondoftwo}%
\providecommand \bibfield  [0]{\@secondoftwo}%
\providecommand \translation [1]{[#1]}%
\providecommand \BibitemOpen [0]{}%
\providecommand \bibitemStop [0]{}%
\providecommand \bibitemNoStop [0]{.\EOS\space}%
\providecommand \EOS [0]{\spacefactor3000\relax}%
\providecommand \BibitemShut  [1]{\csname bibitem#1\endcsname}%
\let\auto@bib@innerbib\@empty
%</preamble>
\bibitem [{\citenamefont {Klempt}\ and\ \citenamefont {Richard}(2010)}]{P1}%
  \BibitemOpen
  \bibfield  {author} {\bibinfo {author} {\bibfnamefont {E.}~\bibnamefont
  {Klempt}}\ and\ \bibinfo {author} {\bibfnamefont {J.~M.}\ \bibnamefont
  {Richard}},\ }\href@noop {} {\bibfield  {journal} {\bibinfo  {journal} {Rev.\
  Mod.\ Phys}\ }\textbf {\bibinfo {volume} {82}},\ \bibinfo {pages} {1095}
  (\bibinfo {year} {2010})}\BibitemShut {NoStop}%
\bibitem [{\citenamefont {Zou}(2001)}]{P2}%
  \BibitemOpen
  \bibfield  {author} {\bibinfo {author} {\bibfnamefont {B.~S.}\ \bibnamefont
  {Zou}},\ }\href@noop {} {\bibfield  {journal} {\bibinfo  {journal} {Nucl.\
  Phys.\ A}\ }\textbf {\bibinfo {volume} {684}},\ \bibinfo {pages} {330}
  (\bibinfo {year} {2001})}\BibitemShut {NoStop}%
\bibitem [{\citenamefont {\relax Ablikim~\textit{et al}.}(2004)}]{P3}%
  \BibitemOpen
  \bibfield  {author} {\bibinfo {author} {\bibfnamefont {M.}~\bibnamefont
  {\relax Ablikim~\textit{et al}.}} (\bibinfo {collaboration} {BES
  Collaboration}),\ }\href@noop {} {\bibfield  {journal} {\bibinfo  {journal}
  {Phys.\ Rev.\ Lett.}\ }\textbf {\bibinfo {volume} {93}},\ \bibinfo {pages}
  {112002} (\bibinfo {year} {2004})}\BibitemShut {NoStop}%
\bibitem [{\citenamefont {\relax Ablikim~\textit{et
  al}.}(2013{\natexlab{a}})}]{P4}%
  \BibitemOpen
  \bibfield  {author} {\bibinfo {author} {\bibfnamefont {M.}~\bibnamefont
  {\relax Ablikim~\textit{et al}.}} (\bibinfo {collaboration} {BESIII
  Collaboration}),\ }\href@noop {} {\bibfield  {journal} {\bibinfo  {journal}
  {Phys.\ Rev.\ D}\ }\textbf {\bibinfo {volume} {87}},\ \bibinfo {pages}
  {012007} (\bibinfo {year} {2013}{\natexlab{a}})}\BibitemShut {NoStop}%
\bibitem [{\citenamefont {\relax Ablikim~\textit{et al}.}(2018)}]{P5}%
  \BibitemOpen
  \bibfield  {author} {\bibinfo {author} {\bibfnamefont {M.}~\bibnamefont
  {\relax Ablikim~\textit{et al}.}} (\bibinfo {collaboration} {BESIII
  Collaboration}),\ }\href@noop {} {\bibfield  {journal} {\bibinfo  {journal}
  {Chin. Phys. C}\ }\textbf {\bibinfo {volume} {42}},\ \bibinfo {pages}
  {023001} (\bibinfo {year} {2018})}\BibitemShut {NoStop}%
\bibitem [{\citenamefont {\relax Ablikim~\textit{et al}.}(2010)}]{P6}%
  \BibitemOpen
  \bibfield  {author} {\bibinfo {author} {\bibfnamefont {M.}~\bibnamefont
  {\relax Ablikim~\textit{et al}.}} (\bibinfo {collaboration} {BESIII
  Collaboration}),\ }\href@noop {} {\bibfield  {journal} {\bibinfo  {journal}
  {Nucl.\ Instrum.\ Methods Phys.\ Res.,\ Sect.\ A}\ }\textbf {\bibinfo
  {volume} {614}},\ \bibinfo {pages} {345} (\bibinfo {year}
  {2010})}\BibitemShut {NoStop}%
\bibitem [{\citenamefont {\relax Agostinelli~\textit{et al}.}(2003)}]{P7}%
  \BibitemOpen
  \bibfield  {author} {\bibinfo {author} {\bibfnamefont {S.}~\bibnamefont
  {\relax Agostinelli~\textit{et al}.}} (\bibinfo {collaboration} {GEANT4
  Collaboration}),\ }\href@noop {} {\bibfield  {journal} {\bibinfo  {journal}
  {Nucl. Instrum. Methods Phys. Res., Sect. A}\ }\textbf {\bibinfo {volume}
  {506}},\ \bibinfo {pages} {250} (\bibinfo {year} {2003})}\BibitemShut
  {NoStop}%
\bibitem [{\citenamefont {Jadach}\ \emph {et~al.}(2001)\citenamefont {Jadach},
  \citenamefont {Ward},\ and\ \citenamefont {Was}}]{P8}%
  \BibitemOpen
  \bibfield  {author} {\bibinfo {author} {\bibfnamefont {S.}~\bibnamefont
  {Jadach}}, \bibinfo {author} {\bibfnamefont {B.}~\bibnamefont {Ward}}, \ and\
  \bibinfo {author} {\bibfnamefont {Z.}~\bibnamefont {Was}},\ }\href@noop {}
  {\bibfield  {journal} {\bibinfo  {journal} {Phys.\ Rev.\ D}\ }\textbf
  {\bibinfo {volume} {63}},\ \bibinfo {pages} {113009} (\bibinfo {year}
  {2001})}\BibitemShut {NoStop}%
\bibitem [{\citenamefont {Lange}(2001)}]{P9}%
  \BibitemOpen
  \bibfield  {author} {\bibinfo {author} {\bibfnamefont {D.~J.}\ \bibnamefont
  {Lange}},\ }\href@noop {} {\bibfield  {journal} {\bibinfo  {journal} {Nucl.\
  Instrum.\ Methods Phys.\ Res.,\ Sect.\ A}\ }\textbf {\bibinfo {volume}
  {462}},\ \bibinfo {pages} {152} (\bibinfo {year} {2001})}\BibitemShut
  {NoStop}%
\bibitem [{\citenamefont {Ping}(2008)}]{P13}%
  \BibitemOpen
  \bibfield  {author} {\bibinfo {author} {\bibfnamefont {R.~G.}\ \bibnamefont
  {Ping}},\ }\href@noop {} {\bibfield  {journal} {\bibinfo  {journal} {Chin.\
  Phys.\ C}\ }\textbf {\bibinfo {volume} {32}},\ \bibinfo {pages} {599}
  (\bibinfo {year} {2008})}\BibitemShut {NoStop}%
\bibitem [{\citenamefont {\relax Patrignani~\textit{et al}.}(2018)}]{P10}%
  \BibitemOpen
  \bibfield  {author} {\bibinfo {author} {\bibfnamefont {C.}~\bibnamefont
  {\relax Patrignani~\textit{et al}.}} (\bibinfo {collaboration} {Particle Data
  Group}),\ }\href@noop {} {\bibfield  {journal} {\bibinfo  {journal} {Phys.\
  Rev.\ D}\ }\textbf {\bibinfo {volume} {98}},\ \bibinfo {pages} {030001}
  (\bibinfo {year} {2018})}\BibitemShut {NoStop}%
\bibitem [{\citenamefont {Chen}\ \emph {et~al.}(2000)\citenamefont {Chen},
  \citenamefont {Huang}, \citenamefont {Qi}, \citenamefont {Zhang},\ and\
  \citenamefont {Zhu}}]{P11}%
  \BibitemOpen
  \bibfield  {author} {\bibinfo {author} {\bibfnamefont {J.~C.}\ \bibnamefont
  {Chen}}, \bibinfo {author} {\bibfnamefont {G.~S.}\ \bibnamefont {Huang}},
  \bibinfo {author} {\bibfnamefont {X.~R.}\ \bibnamefont {Qi}}, \bibinfo
  {author} {\bibfnamefont {D.~H.}\ \bibnamefont {Zhang}}, \ and\ \bibinfo
  {author} {\bibfnamefont {Y.~S.}\ \bibnamefont {Zhu}},\ }\href@noop {}
  {\bibfield  {journal} {\bibinfo  {journal} {Phys.\ Rev.\ D}\ }\textbf
  {\bibinfo {volume} {62}},\ \bibinfo {pages} {034003} (\bibinfo {year}
  {2000})}\BibitemShut {NoStop}%
\bibitem [{\citenamefont {Karl}\ \emph {et~al.}(1976)\citenamefont {Karl},
  \citenamefont {Meshkov},\ and\ \citenamefont {Rosner}}]{P12}%
  \BibitemOpen
  \bibfield  {author} {\bibinfo {author} {\bibfnamefont {G.}~\bibnamefont
  {Karl}}, \bibinfo {author} {\bibfnamefont {S.}~\bibnamefont {Meshkov}}, \
  and\ \bibinfo {author} {\bibfnamefont {J.~L.}\ \bibnamefont {Rosner}},\
  }\href@noop {} {\bibfield  {journal} {\bibinfo  {journal} {Phys.\ Rev.\ D}\
  }\textbf {\bibinfo {volume} {13}},\ \bibinfo {pages} {1203} (\bibinfo {year}
  {1976})}\BibitemShut {NoStop}%
\bibitem [{\citenamefont {\relax Ablikim~\textit{et al}.}(2016)}]{P15}%
  \BibitemOpen
  \bibfield  {author} {\bibinfo {author} {\bibfnamefont {M.}~\bibnamefont
  {\relax Ablikim~\textit{et al}.}} (\bibinfo {collaboration} {BESIII
  Collaboration}),\ }\href@noop {} {\bibfield  {journal} {\bibinfo  {journal}
  {Phys.\ Lett.\ B}\ }\textbf {\bibinfo {volume} {753}},\ \bibinfo {pages}
  {629} (\bibinfo {year} {2016})}\BibitemShut {NoStop}%
\bibitem [{\citenamefont {\relax Ablikim~\textit{et
  al}.}(2013{\natexlab{b}})}]{P16}%
  \BibitemOpen
  \bibfield  {author} {\bibinfo {author} {\bibfnamefont {M.}~\bibnamefont
  {\relax Ablikim~\textit{et al}.}} (\bibinfo {collaboration} {BESIII
  Collaboration}),\ }\href@noop {} {\bibfield  {journal} {\bibinfo  {journal}
  {Phys.\ Rev.\ D}\ }\textbf {\bibinfo {volume} {87}},\ \bibinfo {pages}
  {092006} (\bibinfo {year} {2013}{\natexlab{b}})}\BibitemShut {NoStop}%
\bibitem [{\citenamefont {\relax Anashin~\textit{et al}.}(2011)}]{P17}%
  \BibitemOpen
  \bibfield  {author} {\bibinfo {author} {\bibfnamefont {V.~V.}\ \bibnamefont
  {\relax Anashin~\textit{et al}.}},\ }\href@noop {} {\bibfield  {journal}
  {\bibinfo  {journal} {Int.\ J.\ Mod.\ Phys.\ Conf.\ Ser}\ }\textbf {\bibinfo
  {volume} {02}},\ \bibinfo {pages} {188} (\bibinfo {year} {2011})}\BibitemShut
  {NoStop}%
\bibitem [{\citenamefont {Chung}(1998)}]{P18}%
  \BibitemOpen
  \bibfield  {author} {\bibinfo {author} {\bibfnamefont {S.~U.}\ \bibnamefont
  {Chung}},\ }\href@noop {} {\bibfield  {journal} {\bibinfo  {journal} {Phys.\
  Rev.\ D}\ }\textbf {\bibinfo {volume} {57}},\ \bibinfo {pages} {431}
  (\bibinfo {year} {1998})}\BibitemShut {NoStop}%
\bibitem [{\citenamefont {\relax Chekanov~\textit{et al}.}(2005)}]{P19}%
  \BibitemOpen
  \bibfield  {author} {\bibinfo {author} {\bibfnamefont {S.}~\bibnamefont
  {\relax Chekanov~\textit{et al}.}} (\bibinfo {collaboration} {ZEUS
  Collaboration}),\ }\href@noop {} {\bibfield  {journal} {\bibinfo  {journal}
  {Eur.\ Phys.\ J.\ C}\ }\textbf {\bibinfo {volume} {44}},\ \bibinfo {pages}
  {351} (\bibinfo {year} {2005})}\BibitemShut {NoStop}%
\bibitem [{\citenamefont {\relax Albrecht~\textit{et al}.}(1990)}]{P20}%
  \BibitemOpen
  \bibfield  {author} {\bibinfo {author} {\bibfnamefont {H.}~\bibnamefont
  {\relax Albrecht~\textit{et al}.}} (\bibinfo {collaboration} {ARGUS
  Collaboration}),\ }\href@noop {} {\bibfield  {journal} {\bibinfo  {journal}
  {Phys.\ Lett.\ B}\ }\textbf {\bibinfo {volume} {241}},\ \bibinfo {pages}
  {278} (\bibinfo {year} {1990})}\BibitemShut {NoStop}%
\bibitem [{\citenamefont {\relax Ablikim~\textit{et
  al}.}(2013{\natexlab{c}})}]{P21}%
  \BibitemOpen
  \bibfield  {author} {\bibinfo {author} {\bibfnamefont {M.}~\bibnamefont
  {\relax Ablikim~\textit{et al}.}} (\bibinfo {collaboration} {BESIII
  Collaboration}),\ }\href@noop {} {\bibfield  {journal} {\bibinfo  {journal}
  {Chin.\ Phys.\ C}\ }\textbf {\bibinfo {volume} {37}},\ \bibinfo {pages}
  {123001} (\bibinfo {year} {2013}{\natexlab{c}})}\BibitemShut {NoStop}%
\bibitem [{\citenamefont {\relax Ablikim~\textit{et al}.}(2011)}]{P22}%
  \BibitemOpen
  \bibfield  {author} {\bibinfo {author} {\bibfnamefont {M.}~\bibnamefont
  {\relax Ablikim~\textit{et al}.}} (\bibinfo {collaboration} {BESIII
  Collaboration}),\ }\href@noop {} {\bibfield  {journal} {\bibinfo  {journal}
  {Phys.\ Rev.\ D}\ }\textbf {\bibinfo {volume} {83}},\ \bibinfo {pages}
  {112005} (\bibinfo {year} {2011})}\BibitemShut {NoStop}%
\bibitem [{\citenamefont {\relax Ablikim~\textit{et al}.}(2012)}]{P23}%
  \BibitemOpen
  \bibfield  {author} {\bibinfo {author} {\bibfnamefont {M.}~\bibnamefont
  {\relax Ablikim~\textit{et al}.}} (\bibinfo {collaboration} {BESIII
  Collaboration}),\ }\href@noop {} {\bibfield  {journal} {\bibinfo  {journal}
  {Phys.\ Rev.\ D}\ }\textbf {\bibinfo {volume} {85}},\ \bibinfo {pages}
  {092012} (\bibinfo {year} {2012})}\BibitemShut {NoStop}%
\bibitem [{\citenamefont {\relax Ablikim~\textit{et
  al}.}(2013{\natexlab{d}})}]{P24}%
  \BibitemOpen
  \bibfield  {author} {\bibinfo {author} {\bibfnamefont {M.}~\bibnamefont
  {\relax Ablikim~\textit{et al}.}} (\bibinfo {collaboration} {BESIII
  Collaboration}),\ }\href@noop {} {\bibfield  {journal} {\bibinfo  {journal}
  {Phys.\ Rev.\ D}\ }\textbf {\bibinfo {volume} {87}},\ \bibinfo {pages}
  {012002} (\bibinfo {year} {2013}{\natexlab{d}})}\BibitemShut {NoStop}%
\end{thebibliography}%

\end{document}